\documentclass[prd, twocolumn, nofootinbib]{revtex4-2}

\usepackage{bbm}

\usepackage{amsmath}
\usepackage{amsfonts}
\usepackage{amssymb}

\usepackage{multirow}

\DeclareMathOperator*{\argmin}{argmin}
\DeclareMathOperator*{\argmax}{argmax}

\usepackage{graphicx}
\usepackage{tikz}

\usepackage{color}

\usepackage{comment}

\newcommand{\reed}[1]{\textcolor{red}{#1}}

\newcommand{\result}[1]{#1}

\begin{document}

\title{
Calibration Uncertainty's Impact on Gravitational-Wave Observations
}

\author{Reed Essick}
\affiliation{Perimeter Institute for Theoretical Physics \\ 31 Caroline Street North, Waterloo, ON N2L 2Y5}

\begin{abstract}
    Our ability to characterize astrophysical Gravitational Waves depends on our understanding of the detectors used to observe them.
    Specifically, our ability to calibrate current kilometer-scale interferometers can potentially confound the inference of astrophysical signals.
    Current calibration uncertainties are dominated by systematic errors between the modeled and observed detector response and are well described by a Gaussian process.
    I exploit this description to analytically examine the impact of calibration uncertainty.
    I derive closed-form expressions for the conditioned likelihood of the calibration error given the observed data and an astrophysical signal (astrophysical calibration) as well as for the marginal likelihood for the data given a signal (integrated over the calibration uncertainty).
    I show that calibration uncertainty always reduces search sensitivity and the amount of information available about astrophysical signals.
    Additionally, calibration uncertainty will fundamentally limit the precision to which loud signals can be constrained, a crucial factor when considering the scientific potential of proposed third-generation interferometers.
    For example, I estimate that with $1\%$ uncertainty in the detector response's amplitude and phase, one will only be able to measure the leading-order tidal parameter ($\tilde\Lambda$) for a 1.4+1.4$\,M_\odot$ system to better than \result{$\pm 1$} (\result{$\sim 0.2\%$} relative uncertainty) for signals with \result{signal-to-noise ratios $\gtrsim 10^4$}.
    At this signal-to-noise ratio, calibration uncertainty increases $\sigma_{\tilde\Lambda}$ by a factor of \result{$2$} compared to stationary Gaussian noise alone.
    Furthermore, 1\% calibration uncertainty limits the precision to always be \result{$\sigma_{\tilde\Lambda} \gtrsim 0.5$}.
    I also show how to best select the frequencies at which calibration should be precisely constrained in order to minimize the information lost about astrophysical parameters.
    It is not necessary to constrain the calibration errors to be small at all frequencies to perform precise astrophysical inference for individual signals.
\end{abstract}

\maketitle

\section{Introduction}
\label{sec:introduction}

Modern interferometric Gravitational-Wave (GW) detectors~\cite{LIGO, Virgo, KAGRA} provide unprecedented access to fundamental physics on both the largest and smallest scales.
Advanced detector technologies probe aspects of quantum interactions with macroscopic (classical) objects~\cite{McCuller:2020, Yu:2020}, and the astrophysical signals detected probe gravity in the strong/dynamical-field regime~\cite{GWTC2_TGR, GWTC3_TGR}, the distribution of rare stellar remnants, including black holes (BHs) and neutron stars (NSs)~\cite{GWTC2_RaP, GWTC3_RaP}, the behavior of matter at densities up to several times that of stable nuclei (e.g., Refs.~\cite{Legred:2021, Essick:2021}), and many other extreme environments that are otherwise difficult or impossible to access.
Much of this success can be attributed to the detectors' ability to faithfully report the astrophysical strain so that meaningful comparisons to theoretical expectations can be made.
The process of predicting how astrophysical signals will appear within the detector is referred to as calibration.
It typically proceeds by measuring the opto-mechanical response of the interferometer's cavities and control loops.

Calibration has long been of general interest in the field.
Although many calibration techniques exist, the basic physical measurement is the change in the detector output as a function of the change in the difference between the lengths of the interferometer's arms.
In current interferometric observatories (IFOs), one typically moves the test masses (mirrors) that bracket the Fabry-P\'{e}rot cavity in each arm by a known amount at a known frequency and then measures the detector response.
Initial techniques involved driving the suspension system, composed of a cascade of pendula, that suspend each test mass and isolate it from many sources of terrestrial noise (see, e.g., Refs.~\cite{Viets:2018,Sun:2020} and references therein). 
Recently, several authors have investigated Newtonian calibrators~\cite{Ross:2021, Estevez:2021}, which drive the test masses through the Newtonian gravitational force exerted by an oscillating mass multipoles (rotating dumbbell).
Additionally, several authors have studied the ability of astrophysical signals themselves to calibrate detectors.
In this approach, one assumes that the correct theory of gravity is known and compares the detector output to the expected change in length induced by different astrophysical sources.
Indeed, there is sustained interest in understanding both how to use astrophysical signals to calibrate our detectors and how uncertainty in calibration can affect our inference of the properties astrophysical signals~\cite{Vitale:2011, Pitkin:2015, Essick:2019, Payne:2020, Vitale:2020}.

Currently, the most successful calibration techniques rely on auxiliary laser systems (photon calibrators or PCALs~\cite{Karki:2016, Bhattacharjee:2020}) to exert a known radiation force (laser with known intensity) on the test masses.
PCALs have met with great success, providing precise calibration measurements throughout the advanced detector era~\cite{Cahillane:2017}.
However, through the careful quantification of the detector response enabled by PCAL measurements, it is now understood that systematic modeling uncertainties in the detector response, including variations over time, are typically larger than the uncertainty in the directly-measured parameters of the model for the detector response~\cite{Cahillane:2017, Sun:2020}.
These systematic uncertainties are represented with a Gaussian process (GP) and uncertainty envelopes derived therefrom are available at a high cadence~\cite{O1-O2-O3-hourly-uncertainties}.

It has become common practice to marginalize over the calibration uncertainty at the time of individual events within inferences of those events' astrophysical parameters.
Several techniques exist, and they differ primarily in how they represent the calibration uncertainty.
Traditionally, a cubic spline with knots spaced logarithmically in frequency is used~\cite{Farr:2014}.
The marginal prior distributions for the calibration error at each knot is chosen to match the measured calibration envelope at that frequency.
However, it is not necessarily the case that this spline-prior is a good representation of the full GP, which includes correlations between frequencies and may not be a slowly varying function of frequency.
Several authors have also investigated physical models of calibration uncertainty, even parametrizing the model of the detector response in addition to sampling from the full GP prior for the model systematics~\cite{Payne:2020, Vitale:2020}.
Marginalization over calibration is then performed numerically.
While this approach is computationally tractable, it does not offer an immediate interpretation of which aspects of the calibration uncertainty most affect the astrophysical inference or, in turn, which astrophysical signals could most improve our knowledge of the detectors' calibration.

I show how to analyze GP calibration priors analytically, as the GP prior is conjugate to the GW likelihood under the common assumption of stationary Gaussian noise.
This allows me to derive several expressions for the impact of calibration uncertainty on searches for, and the inference of, astrophysical signals.
I also examine how astrophysical signals can inform our understanding of calibration in our detectors.

While these expressions should be of general interest, current calibration uncertainties are typically small ($O(\mathrm{few}\%)$ in amplitude and phase~\cite{Cahillane:2017, Sun:2020}). 
That is, signals detected with current IFOs are quiet enough that the calibration uncertainty does not matter much~\cite{Lindblom:2009, GWTC1, GWTC2, GWTC2d1, GWTC3}.
This may not be the case for proposed third-generation (3G) detectors, in which the typical signal-to-noise ratio (S/N) may be larger.
I clarify the S/N thresholds above which uncertainty in detector calibration will dominate over uncertainty from stationary Gaussian noise.
What's more, opto-mechanical effects in the detector response for 3G detectors will be much more complicated than in current detectors, as light's round-trip travel time within the Fabry-P\'{e}rot cavities becomes comparable to the frequencies probed in the astrophysical strain~\cite{Essick:2017}.
Altogether, then, astrophysical inference of the loudest events detected with 3G detectors will almost certainly be limited by the understanding of the detector responses, and several common approximations for current detector responses will need to be revisited.
This fact appears to not have been fully investigated within the 3G literature (see, e.g., discussion in Ref.~\cite{CosmicExplorer}), and I hope that the analytic expressions presented in this manuscript will enable improved estimates of the scientific capabilities of 3G instruments.

I first summarize a few basic assumptions common in GW data analysis and introduce my notation in Sec.~\ref{sec:notation}.
I then make use of the GP representation of the calibration uncertainty in Sec.~\ref{sec:marginal likelihood} to obtain analytic expressions for the conditioned distribution for the calibration error given the observed data and an astrophysical signal as well as the marginal distribution for the observed data given an astrophysical signal integrated over uncertainty in detector calibration.
Several limiting cases are explored in Sec.~\ref{sec:limiting cases}, and Sec.~\ref{sec:detector networks and multiple signals} explores extensions to networks of detectors and the joint inference over multiple signals.
I discuss the possible impact on searches in Sec.~\ref{sec:searches and sensitivity}.
Sec.~\ref{sec:fisher matrix} discusses the measurability of astrophysical parameters based on the calibration-marginalized likelihood, with particular focus on how much information about the astrophysical parameters is lost due to uncertainty in the detector response.
The similarity of calibration errors to waveform uncertainty and deviations from General Relativity (GR) is introduced within Sec.~\ref{sec:waveform uncertainty and deviations from general relativity}.
I conclude with a discussion of calibration requirements in Sec.~\ref{sec:conclusion}.

\section{Notation and Stationary Gaussian Noise}
\label{sec:notation}

The likelihood of observing data $d$ in stationary Gaussian noise described by a one-sided power spectral density (PSD) $s$ within a single interferometer in the presence of a signal described by intrinsic parameters $\vartheta$ and extrinsic parameters $\varphi$ is typically written as
\begin{equation}\label{eq:continuous likelihood}
    \ln p(d|\theta, \varphi, \mathcal{H}_{s}) \supset -\frac{1}{2}\left( 4\int\limits_0^\infty \mathrm{d}f\, \frac{|d-h(\vartheta, \varphi)|^2}{s}\right)
\end{equation}
where $h= f_\alpha(\varphi) h_\alpha(\vartheta)$ is the strain projected in the interferometer, $f_\alpha$ is the detector response to the $\alpha$ polarization, $h_\alpha$ is the waveform in the $\alpha$ polarization, and repeated indices imply summation.
In general, the detector response $f_\alpha$ is a function of the frequency~\cite{Essick:2017}.
The PSD is defined in terms of the noise autocorrelation function as
\begin{equation}\label{eq:noise autocorrelation}
    \left< n(t) n(t+\tau)\right> = \frac{1}{2} \int \mathrm{d}f\, e^{2\pi i f \tau} s(f)
\end{equation}
where $i=\sqrt{-1}$ and $\left<\cdot\right>$ denotes an average over noise realizations.
$\mathcal{H}_s$ denotes the assumption of stationary Gaussian noise with a known PSD.
See, e.g. Refs.~\cite{Allen:2012, Talbot:2020} for more discussion about PSD estimation.
Under this assumption, the noise model is a stationary Gaussian process (GP) with (auto)correlations defined by Eq.~\ref{eq:noise autocorrelation}.

However, as IFOs record only discrete data at regularly spaced intervals, it will be convenient to work explicitly with matrices in the following.
In most cases, I suppress indices for simplicity.
However, for notational clarity, I denote objects that have a single frequency index with lower case letters and objects that have two frequency indices with upper case letters.
It will also be convenient to work with only real variables.
As such, I explicitly include the real and imaginary parts of data separately so that
\begin{align}
    h_{2i} & = \mathcal{R}\{h(f_i)\} \\
    h_{2i+1} & = \mathcal{I}\{h(f_i)\}
\end{align}
for $i \in \{1, \cdots, N_\mathrm{frq}$\}.
Therefore, for complex data recorded at $N_\mathrm{frq}$ discrete frequencies, vectors have $2N_\mathrm{frq}$ real entries and matrices have $4N_\mathrm{frq}^2$ real elements.

With this notation, I obtain
\begin{align}
    \ln & p(d| \theta,\varphi, \mathcal{H}_{s}) \nonumber \\
        & = - 2 \sum_i^{N_\mathrm{frq}} \left[ \Delta f\, \frac{\mathcal{R}\{d(f_i) - h(f_i)\}^2 + \mathcal{I}\{d(f_i) - h(f_i)\}^2}{s(f_i)} \right] \nonumber \\
        & \quad\quad\quad\quad\quad\quad - N_\mathrm{frq}\ln 2 \pi - \sum\limits_i^{N_\mathrm{frq}} \ln \left(\frac{s(f_i)}{4\Delta f}\right) \nonumber \\
        & = -\frac{1}{2} (d - h)^\mathrm{T} C^{-1} (d - h) \nonumber \\
        & \quad\quad\quad\quad\quad\quad - N_\mathrm{frq}\ln 2 \pi - \frac{1}{2}\ln \mathrm{det}|C| \label{eq:likelihood}
\end{align}
where $C$ is the covariance matrix (expected to be diagonal) with
\begin{equation}
    C_{2i\,2i} = C_{2i+1\, 2i+1} \equiv \left( \frac{s(f_i)}{4\Delta f} \right)
\end{equation}
for stationary Gaussian noise.
That is, there are neither correlations between frequencies nor between the real and imaginary parts of the data at each frequency.
In general, nonstationary noise and/or windowing functions applied within discrete Fourier transforms may introduce off-diagonal terms within $C$ (see, e.g., Ref.~\cite{Talbot:2021}).
I ignore these for the purposes of this study, although I do not assume $C$ is diagonal in what follows unless explicitly stated.

Now, if the modeled detector response is incorrect by a multiplicative factor such that $f^{(\mathrm{true})}_\alpha = f_\alpha (1 + \delta)$, the actual projected strain in the detector will be $h^{(\mathrm{true})} = h (1 + \delta)$.
This assumes that the calibration uncertainty affects only the total projected strain in the interferometer and therefore impacts the detector response to each polarization in the same proportion.\footnote{This may not be true in general, though, as approximations to the interferometer's response may break down differently for the response to each polarization. These errors may also depend on the source's extrinsic parameters (i.e., relative orientation of the source and the detector). However, these effects are small compared to the GP uncertainty for current detectors, and the approximation that $\delta$ is independent of polarization and $\varphi$ should be reasonable.}
In general, $\delta$ can be complex and mixes the real and imaginary parts of the projected strain so that
\begin{align}
    \mathcal{R}\{h^{(\mathrm{true})}\} & = \mathcal{R}\{h\} + \mathcal{R}\{h\}\mathcal{R}\{\delta\} - \mathcal{I}\{h\}\mathcal{I}\{\delta\} \\
    \mathcal{I}\{h^{(\mathrm{true})}\} & = \mathcal{I}\{h\} + \mathcal{I}\{h\}\mathcal{R}\{\delta\} + \mathcal{R}\{h\}\mathcal{I}\{\delta\}
\end{align}
which I express as
\begin{equation}\label{eq:true strain}
    h^{(\mathrm{true})} = h + H \delta
\end{equation}
where $H$ is a ($2\times2$ block)-diagonal matrix with each block given by
\begin{equation}
    H_{2\times2} = \begin{bmatrix} \mathcal{R}\{h(f_i)\} & - \mathcal{I}\{h(f_i)\} \\ +\mathcal{I}\{h(f_i)\} & \mathcal{R}\{h(f_i)\} \end{bmatrix}
\end{equation}
I can then replace $h$ in Eq.~\ref{eq:likelihood} with Eq.~\ref{eq:true strain} to obtain the likelihood of observing $d$ in the presence of a true signal $h$ and calibration error $\delta$.

It is common to parametrize $\delta$ as
\begin{equation}
    1 + \delta = (1+\delta A)e^{i\delta \psi}
\end{equation}
where both $\delta A$ and $\delta \psi$ are real.
I instead use $\mathcal{R}\{\delta\}$ and $\mathcal{I}\{\delta\}$ because this allows me to marginalize over them analytically.
Nonetheless, note that $\delta \approx \delta A + i \delta \psi$ in the limit $\delta A$, $\delta \psi \ll 1$, which is often the case.
In this limit, then, uncertainty in the amplitude and phase, including correlations between the two, can be immediately translated into uncertainty on the real and imaginary parts of the calibration error.

Often, one has some prior knowledge of $\delta$ (both amplitude and phase) as a function of frequency, which allows one to place a reasonable prior on $\delta$.
This is informed by direct measurements of the detector response and models of the interferometer's cavities~\cite{Cahillane:2017, Sun:2020}.
With such a prior, one can construct a joint distribution for both $d$ and $\delta$ assuming $\delta$ is independent of the noise realization and source properties:
\begin{equation}\label{eq:joint likelihood}
    p(d, \delta|\vartheta, \varphi, \mathcal{H}_{s}, \mathcal{H}_\delta) = p(d|\delta, \vartheta, \varphi, \mathcal{H}_{s}) p(\delta|\mathcal{H}_\delta)
\end{equation}
in which $\mathcal{H}_\delta$ denotes the prior assumptions about the calibration uncertainty.
This as the starting point for Sec.~\ref{sec:marginal likelihood}.

\section{Marginal and Conditioned Likelihoods}
\label{sec:marginal likelihood}

It has become common practice to model the calibration uncertainties with a GP~\cite{Cahillane:2017} in order to capture possible systematic errors between the modeled and measured detector responses.
These systematic errors typically dominate over the uncertainty in the detector response from the measured parameters of the opto-mechanical model~\cite{Sun:2020}, and, as such, I consider a GP prior for $\delta$ with mean $\gamma$ and covariance $\Gamma$ such that
\begin{align}
    \ln p(\delta|\mathcal{H}_\delta)
        & = -\frac{1}{2} (\delta - \gamma)^\mathrm{T} \Gamma^{-1} (\delta - \gamma) \nonumber \\
        & \quad \quad - N_\mathrm{frq}\ln 2\pi - \frac{1}{2} \ln \mathrm{det}|\Gamma|
\end{align}
This choice is particularly convenient because it allows me to analytically marginalize over calibration uncertainties in Eq.~\ref{eq:joint likelihood}.
That is, a GP prior for $\mathcal{R}\{\delta\}$ and $\mathcal{I}\{\delta\}$ is conjugate to the likelihood for stationary Gaussian noise.\footnote{One may be able to approximate more general priors for $\delta$ as sums of GPs.}
To wit, marginalizing over $\delta$ yields
\begin{align}\label{eq:marginal likelihood}
    \ln p(d|\vartheta, \varphi, \mathcal{H}_{s}, \mathcal{H}_\delta) = & \ln p(d|\vartheta, \varphi, \mathcal{H}_{s}) \nonumber \\
        & - \frac{1}{2} \gamma^\mathrm{T} \Gamma^{-1} \gamma - \frac{1}{2} \ln \mathrm{det}|\Gamma| \nonumber \\
        & + \frac{1}{2} \nu^\mathrm{T} A^{-1} \nu + \frac{1}{2} \ln \mathrm{det}|A|
\end{align}
where
\begin{equation}
    \nu = A \left[ H^\mathrm{T} C^{-1} (d-h) + \Gamma^{-1} \gamma \right]
\end{equation}
is a vector that is linear in the data and
\begin{align} \label{eq:A}
    A^{-1} = H^\mathrm{T} C^{-1} H + \Gamma^{-1}
\end{align}
As will be seen in Sec.~\ref{sec:limiting cases}, limiting cases of the prior knowledge about calibration errors determine the behavior of $A$.
Indeed, a comparison of the two terms in $A$ is essentially a question of whether the typical fluctuation in projected strain from calibration errors ($\sim |h|^2 \Gamma$) is large compared to the typical fluctuations from Gaussian noise ($\sim s/4\Delta f$).

Eq.~\ref{eq:marginal likelihood} shows that the marginalization can be expressed as a multiplicative factor that modifies the likelihood obtained by assuming perfect calibration.
In this way, one can quickly reweigh ($\vartheta, \varphi$)-samples drawn assuming perfect calibration under different assumptions about the calibration error \textit{post hoc}.
A similar strategy was investigated via numeric marginalization in Ref.~\cite{Payne:2020}.

Similarly, conditioning on $d$ yields
\begin{align}
    \ln p(\delta|d, \vartheta, \varphi, \mathcal{H}_s, \mathcal{H}_\delta) =
        & -\frac{1}{2} (\delta - \nu)^\mathrm{T} A^{-1} (\delta - \nu) \nonumber \\
        & \quad\quad - N_\mathrm{frq}\ln 2\pi - \frac{1}{2}\ln \mathrm{det}|A| \label{eq:delta | d}
\end{align}
One can approximate the marginal posterior on $\delta$ as a Monte Carlo sum of GPs such that
\begin{equation}
    p(\delta|d, \mathcal{H}_{s}, \mathcal{H}_\delta) \approx \frac{1}{N_\mathrm{smp}}\sum\limits_k^{N_\mathrm{smp}} p(\delta|d,\vartheta^{(k)}, \varphi^{(k)}, \mathcal{H}_{s}, \mathcal{H}_\delta)
\end{equation}
with $N_\mathrm{smp}$ $(\vartheta^{(k)}, \varphi^{(k)})$-samples drawn from Eq.~\ref{eq:marginal likelihood}.
Alternatively, one could draw samples from Eq.~\ref{eq:likelihood} and then construct a sum with nontrivial weights using the correction from Eq.~\ref{eq:marginal likelihood}.

At times, it will be more convenient to express Eq.~\ref{eq:marginal likelihood} in terms of a single contraction involving $d$.
This allows me to investigate the effective covariance matrix for $d$ induced by the marginalization over calibration uncertainty.
I therefore write
\begin{align}
    \ln p(d|\vartheta, \varphi, \mathcal{H}_s, \mathcal{H}_\delta) & = -\frac{1}{2} (d - h - \mu)^\mathrm{T} B^{-1} (d - h - \mu) \nonumber \\
        & \quad\quad - N_\mathrm{frq}\ln 2\pi - \frac{1}{2}\ln \mathrm{det}|B| \label{eq:alternate marginal likelihood} \\
\end{align}
with effective (inverse) covariance matrix
\begin{equation}
    B^{-1} = C^{-1} - C^{-1} H A H^\mathrm{T} C^{-1}
\end{equation}
and
\begin{equation}
    \mu = B C^{-1} H A \Gamma^{-1} \gamma
\end{equation}
These are my basic results, and I explore their applications in what follows.

\subsection{Limiting Cases}
\label{sec:limiting cases}

I now consider a few limiting cases: when the changes in the observed data due to calibration uncertainties are small (Sec.~\ref{sec:small deviations}) or large (Sec.~\ref{sec:big deviations}) and when the correlations between the calibration error at neighboring frequencies are small (Sec.~\ref{sec:small correlations}) or large (Sec.~\ref{sec:big correlations}).
Throughout, the following approximation is useful
\begin{equation}
    \left(A + X\right)^{-1} \approx A^{-1} - A^{-1} X A^{-1} + A^{-1} X A^{-1} X A^{-1} + \cdots
\end{equation}
which holds when $||X|| \ll ||A||$.

\subsubsection{Small Deviations: $H \Gamma H^\mathrm{T} \ll C$}
\label{sec:small deviations}

I first consider the case where the calibration uncertainty is small compared to the signal amplitude.
That is, I consider the case where $|h|^2 \Gamma \ll s/4\Delta f$, or the change in the projected strain from calibration errors is small compared to the typical noise fluctuation.
This is the case for current detectors~\cite{Cahillane:2017, Sun:2020}.

When the uncertainty in the calibration is small \textit{a priori},
\begin{align}
    \nu & \approx \gamma + \Gamma H^\mathrm{T} C^{-1} (d-h) \\
    A   & \approx \Gamma - \Gamma H^\mathrm{T} C^{-1} H \Gamma
\end{align}
describing the conditioned uncertainty for $\delta \, | \, d, h$.
While the leading-order change to $A$ is small, note that it acts to decrease the uncertainty in $\delta$.
As such, one gains a small amount of information about the calibration in the presence of a signal even when the detectors are well calibrated \textit{a priori}.

Furthermore,
\begin{align}
    \mu & \approx H \gamma - H \left[ \Gamma H^\mathrm{T} C^{-1} H \right]^2  \gamma \\
    B^{-1} & \approx C^{-1}  -  C^{-1} H \Gamma H^\mathrm{T} C^{-1}
\end{align}
so that
\begin{equation}
    B \approx C + H \Gamma H^\mathrm{T}
\end{equation}
describing the likelihood marginalized over calibration uncertainty.
While the correction in the effective covariance matrix is small (does not strongly affect the astrophysical inference~\cite{Essick:2019, Payne:2020, Vitale:2020}), note that it tends to increase the uncertainty by an amount that is proportional to the signal's amplitude squared.
Although this does not hold in general, one can interpret the effective covariance as the sum of the variance from stationary Gaussian noise and independent variance from calibration uncertainty.

These results make sense intuitively, as the calibration errors are constrained to be small \textit{a priori} and do not add much additional model freedom compared to an analysis that assumes no calibration error.

\subsubsection{Large Deviations: $C \ll H \Gamma H^\mathrm{T}$}
\label{sec:big deviations}

When one does not have strong prior limits on the calibration error in comparison to the projected signal strength in the detector, one does not know how the signal will actually appear.
That is, the uncertainty in the projected strain is large compared to typical noise fluctuations: $|h|^2 \Gamma \gg s/4\Delta f$.
One therefore expects only weak limits on the astrophysical signal.
Specifically,
\begin{align}
    \nu & \approx H^{-1} (d-h) + \left[ \Gamma H^\mathrm{T} C^{-1} H \right]^{-1} \gamma \\
    A   & \approx \left( H^\mathrm{T} C^{-1} H \right)^{-1} - \left[\Gamma H^\mathrm{T} C^{-1} H\right]^{-1} \left( H^\mathrm{T} C^{-1} H \right)^{-1}
\end{align}
and
\begin{align}
    \mu & = H \gamma - H \left[ \Gamma H^\mathrm{T} C^{-1} H \right]^{-2} \gamma \\
    B & \approx H \Gamma H^\mathrm{T} + C
\end{align}
Note that, again, to leading order $B$ is the sum of two sources of variance.
However, in this limit, $B \approx H \Gamma H^\mathrm{T} \gg C$, and the effective covariance matrix is much broader than that of stationary Gaussian noise.

This also means that the uncertainty in the calibration given an astrophysical signal scales inversely with the size of the signal (covariance of $\delta\, |\, d, h$ is $A \sim s/|h|^2$), and one loses most of their ability to constrain the astrophysical signal (covariance of $d\,|\,h$ is $B \sim |h|^2 \Gamma$).
Again, this makes sense.
This limit \textit{de facto} means that one does not know how the astrophysical signal will appear within the detector and therefore cannot constrain it based on the observed data.
Indeed, note that
\begin{equation}
    \ln p(d|\vartheta, \varphi, \mathcal{H}_s, \mathcal{H}_\delta) \sim -\frac{1}{2}(d-h)(H^\mathrm{T})^{-1} \Gamma^{-1} H^{-1} (d-h)
\end{equation}
and the product of $H$ and $(d-h)$ in each $2\times2$ block becomes
\begin{widetext}
\begin{equation}
    H^{-1} (d - h) = \frac{1}{\mathcal{R}\{h\}^2 + \mathcal{I}\{h\}^2} \begin{bmatrix} \mathcal{R}\{h\}\mathcal{R}\{d-h\} + \mathcal{I}\{h\}\mathcal{I}\{d-h\} \\ \mathcal{R}\{h\}\mathcal{I}\{d-h\} - \mathcal{I}\{h\}\mathcal{R}\{d-h\} \end{bmatrix}
\end{equation}
\end{widetext}
which simplifies to $h^\ast (d-h)/|h|^2 = (d/h) -1$, where $d$ and $h$ are complex scalars (rather than vectors of real numbers) and $(\cdot)^\ast$ denotes complex conjugation.
As such, with a slight abuse of notation,
\begin{equation}
    \ln p(d|\vartheta, \varphi, \mathcal{H}_s, \mathcal{H}_\delta) \sim -\frac{1}{2} \left(\frac{d}{h} -1\right)^\mathrm{T} \Gamma^{-1} \left( \frac{d}{h} - 1 \right)
\end{equation}
The effective comparison performed by the inference, then, is simply how well the data reproduces the expected signal and how likely the deviations are according to the calibration prior.


\subsubsection{Small Correlations between frequencies}
\label{sec:small correlations}

When correlations between frequencies are small in both $C$ and $\Gamma$, the likelihood decomposes into $2\times2$ blocks.
The relevant expressions become sums over frequency, with each summand a contraction of vectors of the real and imaginary data at that frequency.
If one makes no assumptions about $\Gamma$ (i.e., real and imaginary parts of $\delta$ may be correlated \textit{a priori}), then it is difficult to make more progress.
However, if one additionally assumes the real and imaginary calibration errors are uncorrelated at each frequency ($\Gamma$ is diagonal), then for each $2\times2$ block 
\begin{equation}
    A_{2\times2} = \begin{bmatrix} \left(\frac{4\Delta f |h|^2}{s} + \frac{1}{\Gamma_{\mathcal{R}\mathcal{R}}} \right)^{-1} & 0 \\ 0 & \left(\frac{4\Delta f |h|^2}{s} + \frac{1}{\Gamma_{\mathcal{I}\mathcal{I}}} \right)^{-1} \end{bmatrix}
\end{equation}
where
\begin{equation}
    \Gamma_{2\times2} = \begin{bmatrix} \Gamma_{\mathcal{R}\mathcal{R}} & 0 \\ 0 & \Gamma_{\mathcal{I}\mathcal{I}} \end{bmatrix}
\end{equation}
Similarly, the effective covariance matrix becomes
\begin{widetext}
\begin{equation}\label{eq:2x2 effective covariance}
    B_{2\times2} \propto 
        \begin{bmatrix} 
            \frac{s}{4\Delta f} + \mathcal{R}\{h\}^2 \Gamma_{\mathcal{R}\mathcal{R}} + \mathcal{I}\{h\}^2\Gamma_{\mathcal{I}\mathcal{I}} & \Gamma_{\mathcal{I}\mathcal{I}} - \Gamma_{\mathcal{R}\mathcal{R}} \\ 
            \Gamma_{\mathcal{I}\mathcal{I}} - \Gamma_{\mathcal{R}\mathcal{R}} & \frac{s}{4\Delta f} + \mathcal{R}\{h\}^2 \Gamma_{\mathcal{I}\mathcal{I}} + \mathcal{I}\{h\}^2\Gamma_{\mathcal{R}\mathcal{R}}
        \end{bmatrix}
\end{equation}
\end{widetext}
If one further specializes to the case where the uncertainty on the real and imaginary parts of the calibration error are equal ($\Gamma_{\mathcal{R}\mathcal{R}} = \Gamma_{\mathcal{I}\mathcal{I}} = \sigma_\Gamma^2$), the coefficient of proportionality in Eq.~\ref{eq:2x2 effective covariance} is unity and $B$ is diagonal with
\begin{equation}
    B_{2\times2} = \left( \frac{s}{4\Delta f} + |h|^2 \sigma_\Gamma^2 \right) \mathbbm{1}_{2\times2}
\end{equation}
The calibration uncertainty acts to produce extra noise in the detector at each frequency, and that extra noise scales with the size of the signal at that frequency.
Note that there is no limit to how much the calibration uncertainty can hurt the inference: $B \rightarrow \infty$ as $\sigma_\Gamma \rightarrow \infty$.

\subsubsection{Large Correlations between frequencies}
\label{sec:big correlations}

When there are large correlations between the calibration errors at different frequencies \textit{a priori}, $\Gamma$ may be rank deficient and therefore may not have an inverse.
As such, many of our expressions can become unwieldy as they depend on $\Gamma^{-1}$.
In the limit of perfect correlations between frequencies (i.e., frequency-independent calibration errors), this becomes particularly troublesome.
This could correspond, for example, to a global calibration amplitude uncertainty induced by errors in the gold-standard integrating sphere~\cite{Tuyenbayev:2020, Lecoeuche:2020, Karki:2016, Bhattacharjee:2020} used to calibrate all other PCALs in the LIGO detectors.
This would be perfectly degenerate with the luminosity distance to sources and could fundamentally limit standard siren cosmology~\cite{Pitkin:2015, Essick:2019}.

I investigate this limit by re-expressing the joint distribution (Eq.~\ref{eq:joint likelihood}) in terms of a direct sum over frequencies (assuming $C$ is diagonal).
To wit, 
\begin{widetext}
\begin{align}
    \ln p(d, \delta| \vartheta, \varphi, \mathcal{H}_s, \mathcal{H}_\delta)
       & \supset -\frac{1}{2} \sum_f (d_f - h_f - H_f \delta)^\mathrm{T} C^{-1}_{ff} (d_f - h_f - H_f \delta) - \frac{1}{2} (\delta - \gamma)^\mathrm{T} \Gamma^{-1} (\delta - \gamma) \nonumber \\
       & \supset -\frac{1}{2} \sum_f (d_f-h_f)^\mathrm{T} C^{-1}_{ff} (d_f-h_f) - \frac{1}{2} \gamma^\mathrm{T} \Gamma^{-1} \gamma \nonumber \\
       & \quad\quad\quad\quad - \frac{1}{2} \delta^\mathrm{T} \left[\Gamma^{-1} + \sum_f H^\mathrm{T}_f C^{-1}_{ff} H_f \right] \delta + \left[ \gamma^\mathrm{T} \Gamma^{-1} + \sum_f (d_f-h_f)^\mathrm{T} C^{-1}_{ff} H_f \right] \delta
\end{align}
\end{widetext}
This is equivalent to the previous expressions except that $\delta$ and $\gamma$ are vectors with only 2 elements, $\Gamma$ is a $2\times2$ matrix, and $H$ is a $2N_\mathrm{frq}\times2$ matrix.
From this, one can immediately write down the terms analogous to the frequency-dependent expressions.
That is, the inverse-covariance between $\mathcal{R}\{\delta\}$ and $\mathcal{I}\{\delta\}$ is a $2\times2$ matrix
\begin{align}\label{eq:large correlations A}
    A^{-1}_{2\times2} & = \Gamma^{-1}_{2\times2} + H^\mathrm{T} C^{-1} H \nonumber \\
           & = \Gamma^{-1}_{2\times2} + \rho^2 \mathbbm{1}_{2\times2}
\end{align}
where\footnote{Eq.~\ref{eq:optimal snr} and the second line in Eq.~\ref{eq:large correlations A} assume $C$ is diagonal.} $\rho$ is the optimal S/N for stationary Gaussian noise, defined as
\begin{equation} \label{eq:optimal snr}
    \rho^2 \equiv h^\mathrm{T} C^{-1} h = 4 \int df \frac{|h|^2}{s}
\end{equation}
Note that the calibration uncertainty can be directly compared to the S/N.
That is, calibration uncertainties only matter when $\Gamma \gtrsim 1/\rho^2$.

The marginal likelihood for $d\,|\,h$ follows the same form as before with
\begin{equation} \label{eq:large correlations C}
    B^{-1} = C^{-1} - C^{-1} H A H^\mathrm{T} C^{-1}
\end{equation}
where, again, $H$ is a $2N_\mathrm{frq}\times2$ matrix and $A$ is a $2\times2$ matrix.
There are potentially large correlations between frequencies in $B$ (the second term in Eq.~\ref{eq:large correlations C} contains the product of ($\mathcal{R}\{h\}$, $\mathcal{I}\{h\}$) at different frequencies).
That is, the strong correlations between $\delta$ at different frequencies induce strong correlations between $d$ at different frequencies.
One can think of this as due to the fact that data at different frequencies witness the same calibration error.

Also note that, when $\Gamma_{2\times2} = \sigma_\Gamma^2 \mathbbm{1}_{2\times2}$
\begin{equation}
    A_{2\times2} = \frac{\sigma^2_\Gamma}{1 + \sigma^2_\Gamma \rho^2} \mathbbm{1}_{2\times2}
\end{equation}
There is a limit to the size of the second term in Eq.~\ref{eq:large correlations C} as $\sigma_\Gamma \rightarrow \infty$.
What's more, the overall correction scales as $H A H^\mathrm{T} C^{-1} \sim A \rho^2$ so that the size of the change in B is independent of both $\sigma_\Gamma$ and $\rho$ in the limit where their product is large.
That is, even horrific frequency-independent calibration uncertainty only modifies the marginal likelihood by a fixed amount.
Therefore, the inference of very loud signals still improves as the signal strength grows ($B \sim \mathrm{constant}$ and $\ln p \sim |h|^2$).
This is in contrast to independent calibration uncertainty at each frequency (Sec.~\ref{sec:small correlations}, $B \propto |h|^2$ and $\ln p \rightarrow \mathrm{constant}$ as $\sigma_\Gamma$ diverges).

\subsection{Detector Networks and Multiple Signals}
\label{sec:detector networks and multiple signals}

So far, I have considered a single astrophysical signal in a single detector.
However, it is straightforward to extend these results to networks of multiple detectors or the simultaneous analysis of multiple signals.
Specifically, one can construct a vector of all the data recorded for each event in each detector as
\begin{align}
    d & = \left. \left. \bigotimes_A^{N_\mathrm{evn} N_\mathrm{IFO}} \right[ d_A \right] \in \mathcal{R}^{2N_\mathrm{frq} N_\mathrm{evn} N_\mathrm{IFO}}
\end{align}
with analogous extensions for the signal models (separate astrophysical parameters for each event) and calibration error.
The subscript $A$ denotes the combination of detector and event.
In a similar way, it is straightforward to generalize, e.g., the prior covariance matrix for calibration errors
\begin{equation}
    \Gamma \rightarrow \begin{bmatrix} \Gamma_{AA} & \cdots & \Gamma_{AN} \\ \vdots & \ddots & \vdots \\ \Gamma_{NA} & \cdots & \Gamma_{NN} \end{bmatrix}
\end{equation}
with the understanding that $\Gamma_{AN} = (\Gamma_{NA})^\mathrm{T}$.
This encodes correlations between calibration errors in different detectors and/or between different events (e.g., calibration errors in each detector may be correlated over time).
Under the assumption of stationary Gaussian noise that is uncorrelated between detectors, $C_{AB}$ is completely diagonal:
\begin{equation}
    C \rightarrow \begin{bmatrix} C_{AA} & & 0 \\  & \ddots &  \\ 0 & & C_{NN} \end{bmatrix}
\end{equation}

With this notation, the corresponding expression for the conditioned and marginal likelihoods are no more complex than for a single IFO.
In particular, note that the special case of independent calibration errors in each detector for a single event ($\Gamma_{AB}$ is block-diagonal) can be handled similarly to the case of calibration errors that are uncorrelated between frequencies.
The likelihood becomes a product of separate single-IFO terms.
Similarly, perfectly correlated calibration errors in each detector can be handled like the case of perfect correlations between frequencies; one can express the likelihood as a direct sum over detectors and obtain expressions like Eqs.~\ref{eq:large correlations A} and~\ref{eq:large correlations C}.
One can see, then, that the multi-IFO analog of Eq.~\ref{eq:large correlations A} implies that one can learn more about (perfectly correlated) calibration errors with a network of detectors or multiple events as the covariance of $\delta \, | \, d, h$ depends on terms like the network S/N (summed in quadrature over events) instead of the S/N in a single IFO (for a single event).

Of particular importance is the case of multiple signals observed with the same detector network.
One is faced with a choice about how to structure the prior for $\delta$.
That is, if one believes the calibration errors are constant over long periods of time, then one may believe that there are strong correlations between events.
Each event sees the same $\delta$, and one can attempt to constrain that $\delta$ directly.
In this case, one computes a posterior for $\delta$ with a single GP prior, obtaining analytic conditional distribution for $\delta \, | \, \{d_A, h_A\}$.

However, if one believes $\delta$ changes more rapidly than the time between events but nevertheless is drawn from a consistent distribution, then one can try to constrain the distribution of $\delta$ based on the observed catalog of events.
In this case, it is the prior for $\delta$ that is common between events, not $\delta$ itself.
Instead of constraining just $\delta$ for each event, one constrains $\gamma$ and $\Gamma$.
That is, one can infer the GP from which a realization of $\delta$ is drawn for each event in a hierarchical inference scheme.
Furthermore, the normal-inverse-Wishart distribution is a conjugate prior for a multivariate Gaussian with unknown mean and covariance.
If one assumes a normal-inverse-Wishart hyperprior for $\gamma$ and $\Gamma$, the entire hierarchical inference remains analytic and can be expressed in a closed form.
That is, one can obtain a posterior distribution for $\gamma, \Gamma \, | \, \{d_A, h_A\}$ analytically.

The differences between these two types of assumptions are common in, e.g., parametrized tests of GR and have been discussed in some detail within that context.
I refer the reader to the excellent reviews in Refs.~\cite{Zimmerman:2019, Isi:2019} for more discussion.

\section{Searches and Sensitivity}
\label{sec:searches and sensitivity}

A search for a known signal morphology that assumes perfect calibration produces a maximum likelihood estimator for the signal amplitude (denoting the signal $h=a\tilde{h}$)
\begin{align}
    \hat{a}_{\sigma_\Gamma=0}
        & = \frac{d^\mathrm{T} C^{-1} \tilde{h}}{\tilde{h}^\mathrm{T} C^{-1} \tilde{h}} \nonumber \\
        & = \left( 4\int df \frac{\mathcal{R}\{d^\ast \tilde{h}\}}{s} \right) \left( 4\int df\, \frac{|\tilde{h}|^2}{s} \right)^{-1}
\end{align}
which is an unbiased estimator with unit variance under the common convention $\tilde{h}^\mathrm{T} C^{-1} \tilde{h} = 1$.
When $\tilde{h}$ is normalized in this way, the estimator is often written as
\begin{equation}\label{eq:matched-filter snr}
    \hat{a}_{\sigma_\Gamma=0} = \frac{d^\mathrm{T} C^{-1} \tilde{h}}{\sqrt{\tilde{h}^\mathrm{T} C^{-1} \tilde{h}}}
\end{equation}
This is the normal definition of the matched-filter S/N.
I examine the behavior of this statistic in the presence of calibration uncertainty.

Specifically, when $\gamma=0$,
\begin{align}
    E[\hat{a}_{\sigma_\Gamma=0}]
        & = a_\mathrm{true} \label{eq:snr mean} \\
    \mathrm{Var}[\hat{a}_{\sigma_\Gamma = 0}]
        & = E[(\hat{a}_{\sigma_\Gamma=0} - E[\hat{a}_{\sigma_\Gamma = 0}])^2] \nonumber \\
        & = \frac{\tilde{h}^\mathrm{T} C^{-1} B C^{-1} \tilde{h}}{\tilde{h}^\mathrm{T} C^{-1} \tilde{h}} > 1 \label{eq:snr variance}
\end{align}
That is, the mean S/N is not affected for zero-mean calibration priors, but the variance is larger ($B \geq C$).
That is, I assume $C$ is the covariance of $d$ in the absence of a signal.
I assume this is known (measured) and is independnet of $\delta$ (Eq.~\ref{eq:likelihood}).\footnote{Ref.~\cite{Lindblom:2008}, instead, suggests the measured $C$ will depend on $\delta$. This means one must include $C$'s dependence on $\delta$ when computing moments of $\hat{a}_{\sigma_\Gamma=0}$. They find a cancellation between the numerator and denominator of Eq.~\ref{eq:matched-filter snr} at leading order in $\gamma$ when computing the expected value.
In the context of their analysis, my result would hold if one performed the inference directly on (any deterministic function of) the raw detector output. Nonetheless, although not shown explicitly, their analysis suggests $\mathrm{Var}[\hat{a}_{\sigma_\Gamma=0}] \geq 1$ as well, which is my main result.}
This implies that, for a given observed S/N, the probability that noise alone (combined with calibration uncertainty) could have produced that statistic is higher.
Therefore, the statistical significance of that S/N is lower.
If one considers the the ratio of the estimator to its standard deviation as a measure of significance (a classical $z$-test), then $\hat{a}_\mathrm{naive}$ will produce strictly smaller $z$ than one would expect without calibration uncertainty.
This makes sense intuitively; it should be harder to detect signal with a matched filter in the presence of calibration uncertainty.

Similar effects occur for unmodeled searches.
The variance of the naive search statistic (maximized likelihood) is increased under the noise hypothesis in the presence of calibration uncertainty.
Louder signals are needed to stand out against the background.

It is noteworthy that modern GW searches do not estimate the false alarm probability based on the assumption of stationary Gaussian noise.
Instead, they employ a variety of bootstrap procedures (see, e.g., Refs.~\cite{Was:2009, Cannon:2013, Usman:2016, Lynch:2017, Drago:2021, Aubin:2021}).
\textit{De facto}, then, there is some threshold above which the observed S/N must fall to be considered a confident detection.
In this context, the broader variance of the naive S/N statistic in the presence of nonzero calibration uncertainty still hampers detectability.
This is because the true S/N (sometimes called the optimal S/N, defined in Eq.~\ref{eq:optimal snr}) must be systematically larger than the threshold in order for the observed S/N to be above the threshold with high probability.
That is, the optimal S/N must be shifted above the threshold by a larger amount when $\Gamma$ is nonzero because the variance of S/N is larger.
Larger optimal S/N imply closer sources, and therefore the space-time volume to which detectors are sensitive is smaller.

Finally, one could derive a maximum likelihood estimator for the signal amplitude directly from $\ln p(d|\vartheta,\varphi, \mathcal{H}_s, \mathcal{H}_\delta)$.
However, the corresponding expressions quickly become intractable and may be computationally prohibitive.
As such, it is expected that they will be of little practical use.

\section{Calibration-Marginalized Fisher Information Matrix}
\label{sec:fisher matrix}

In Sec.~\ref{sec:marginal likelihood}, I obtained general expressions for the marginal likelihood given prior uncertainty on the detector calibration.
I now investigate how the presence of imperfect calibration uncertainty can confound the inference of astrophysical properties in more detail.
Specifically, I consider the Fisher information matrix for astrophysical parameters derived from the calibration-marginalized likelihood.
In particular, I contrast this to what one might naively expect under the assumption that calibration errors were known to be absent.

The Fisher information matrix element between parameters $\vartheta_a$ and $\vartheta_b$ is defined as
\begin{equation}
    I_{ab} \equiv \int \mathcal{D} d \, p(d|\vartheta) \left( \frac{\partial \ln p(d|\vartheta)}{\partial \vartheta_a} \right) \left(\frac{\partial \ln p(d|\vartheta)}{\partial \vartheta_b}\right)
\end{equation}
and is inversely related to the lower bound on the covariance between these variables (Cram\'er-Rao bound).
That is, larger $I_{ab}$ imply smaller covariances and \textit{vice versa}.
In the special case where $p(d|\vartheta)$ is Gaussian
\begin{multline}
    \ln p(d|\vartheta) \supset - \frac{1}{2} (d - h - \mu)^\mathrm{T} B^{-1} (d - h - \mu) \\ - \frac{1}{2}\ln \mathrm{det}|B|
\end{multline}
the Fisher information matrix becomes
\begin{multline} \label{eq:multivariate Gaussian fisher information}
    I_{ab} = \left(\frac{\partial (h + \mu)}{\partial \vartheta_a}\right)^\mathrm{T} B^{-1} \left(\frac{\partial (h + \mu) }{\partial \vartheta_b} \right) \\ + \frac{1}{2}\mathrm{tr}\left[B^{-1} \frac{\partial B}{\partial \vartheta_a} B^{-1} \frac{\partial B}{\partial \vartheta_b} \right]
\end{multline}
The general expression can be quite complicated, and I do not attempt to describe it in detail here.
Instead, I again focus on limiting cases: when the calibration uncertainty is small and when it is large.
In all cases, I assume zero-mean prior processes ($\gamma = \mu = 0$) for simplicity.

For reference, in the naive case ($\Gamma=0$)
\begin{align}
    I^{(\mathrm{naive})}_{ab} & = (\partial_a h) C^{-1} (\partial_b h) \nonumber \\
                              & = 4 \int df \frac{\mathcal{R}\{(\partial_a h)^\ast (\partial_b h)\}}{s}
\end{align}

\subsection{Calibration uncertainties are small}
\label{sec:fisher with small cal uncertainty}

When calibration uncertainties are small and the mean of the prior process is zero,
\begin{multline}
    I_{ab} \approx (\partial_a h)^\mathrm{T} \left( C^{-1} - C^{-1} H \Gamma H^\mathrm{T} C^{-1} \right) (\partial_b h) \\ + \frac{1}{2}\mathrm{tr}\left[C^{-1} (\partial_a (H \Gamma H^\mathrm{T})) C^{-1} (\partial_b (H \Gamma H^\mathrm{T})) \right]
\end{multline}
The leading-order correction (in $\Gamma$) to the naive result is
\begin{align}
    \Delta I_{ab}
        & \equiv I_{ab} - I^{(\mathrm{naive})}_{ab} \nonumber \\
        & = - (\partial_a h)^\mathrm{T} C^{-1} H \Gamma H^\mathrm{T} C^{-1} (\partial_b h)
\end{align}
This term is negative semidefinite when $a=b$ because $\Gamma$ is positive semidefinite.
As such, $I_{aa}$ is always smaller than the naive estimate, implying the variance of $\vartheta_a$ will always be larger than in the naive estimate.
This makes sense.
Imperfect knowledge of the detector's calibration should always worsen the precision to which one can measure astrophysical parameters.

This extra uncertainty is just an inner product of the derivatives of the signal model.
Further note that, assuming $C$ is diagonal, one can write the $2\times2$ blocks
\begin{equation}
    H^\mathrm{T} C^{-1} \partial_b h = \frac{4\Delta f}{s} \begin{bmatrix} \mathcal{R}\{h\}\partial_a \mathcal{R}\{h\} + \mathcal{I}\{h\}\partial_a \mathcal{I}\{h\} \\ \mathcal{I}\{h\}\partial_a \mathcal{R}\{h\} - \mathcal{R}\{h\}\partial_a \mathcal{I}\{h\} \end{bmatrix}
\end{equation}
which is simply $4\Delta f (h^\ast \partial_a h) / s$ (again, reverting to complex scalars instead of real vectors).
The change in $I_{aa}$ is simply the projection of $(h^\ast \partial_a h / s)$ onto the principle components of $\Gamma$.
In this way, $\Gamma$ with large Frobenius norm and/or large marginal uncertainty at a particular frequency may still correspond to small changes in the measurability of certain astrophysical parameters because of the correlations between frequencies.
That is, $h^\ast \partial_a h / s$ may match eigenvectors of $\Gamma$ with small eigenvalues.
However, if the Frobenius norm is small, then $|\Delta I_{ab}|$ will also be small.
Therefore, a small Frobenius norm is a sufficient condition for precise inference, but it may not be necessary.

One may also ask, given the current state of uncertainty, which frequencies should be measured more precisely to minimize the amount of information lost to calibration uncertainty within the astrophysical inference.
That is, minimize $|\Delta I_{ab}|$ by additionally constraining $\delta$ at a single frequency.
If one measures $\delta$ precisely at a single frequency, then the calibration uncertainty is modified so that $\Gamma \rightarrow \Gamma^\prime(f) = (\Gamma^{-1} + Z^{-1}(f))^{-1}$ where $Z$ is diagonal with all matrix elements equal to zero except for at one frequency, where $Z$ is small.
That is, $Z$ represents the covariance from the measurement of $\delta$ at a single frequency.

Using this updated uncertainty, one can then pose the question as follows:
\begin{equation}\label{eq:optimization of calibration measurements}
    f_\mathrm{opt} = \argmin_f \left\{ \left(\partial_a h\right)^\mathrm{T} C^{-1} H \, \Gamma^\prime(f) \, H^\mathrm{T} C^{-1} \partial_b h \right\}
\end{equation}
In general, it is difficult to make analytic progress with this expression.
However, it should be relatively simple to implement numerically.
That is, one can perform a direct search over frequencies, simulating the effect of a precise measurement for each frequency by inserting the appropriate $Z(f)$.
Indeed, this numeric procedure can be extended to an arbitrary number of hypothetical calibration measurements.
One can simply define $Z(f_i)$ for each hypothetical measurement and include the sum within Eq.~\ref{eq:optimization of calibration measurements}.
While the computational cost of a direct search grows quickly with the number of frequency measurements, in practice this may be limited to only a handful of frequencies. 

As an example, consider diagonal $C$ and $\Gamma$.
In this case
\begin{equation}
    |\Delta I_{aa}| = \sum\limits_f \frac{|h_f^\ast \partial_a h_f|^2}{C_f^2} \left(\frac{Z_f \Gamma_f}{Z_f + \Gamma_f}\right)
\end{equation}
Now, in the neighborhood of the new (precise) observation, $Z \rightarrow 0$.
Therefore, precise observations of $\delta$ act as a filter that zeros out that neighborhood within the integral that determines $|\Delta I_{aa}|$.
By inspection, then, assuming $\Gamma$ is the same at all frequencies
\begin{equation}
    f_\mathrm{opt} = \argmax_f \left\{ \left(\frac{|h|^2}{s}\right)\left(\frac{|\partial_a h|^2}{s}\right) \right\}
\end{equation}
That is, one should measure the calibration at the frequency at which the current calibration uncertainty most hurts the inference.
This is a combination of the rate at which information is gained ($|\partial_a h|^2/s$) and when the signal (and therefore calibration uncertainty's impact on the observed data) is large compared to Gaussian noise ($|h|^2/s$).
However, when there are correlations between different frequencies encoded within $\Gamma$, the selection of $f_\mathrm{opt}$ may be more complicated.

Note that, in general, $f_\mathrm{opt}$ will depend on both $h$ and the parameter of interest.
A broader optimization procedure should consider the ability to constrain multiple astrophysical parameters simultaneously for many different possible signals.
For example, it may be most advantageous to constrain $\delta$ at relatively high frequencies in order to minimize calibration uncertainty's impact on the measurement of tidal deformability in binary neutron star (BNS) mergers.
However, those frequencies may be above the maximum frequency produced in binary black hole (BBH) mergers.
Additional calibration measurements at high frequencies, which most benefit BNSs, may not improve astrophysical inference for BBHs (barring strong inter-frequency correlations in $\Gamma$).
As such, while a small Frobenius norm for $\Gamma$ is not necessary for small $|\Delta I_{ab}|$ for any individual signal, it may be the simplest way to guarantee small $|\Delta I_{ab}|$ for all signals.

\subsection{Calibration uncertainties are large}
\label{sec:fisher with large cal uncertainty}

In the limit of large calibration uncertainties and $\gamma=0$, $B \sim H \Gamma H^\mathrm{T} \sim |h|^2 \Gamma$.
Eq.~\ref{eq:multivariate Gaussian fisher information} immediately shows that the uncertainty in $\vartheta$ does not depend on the signal amplitude.
That is, louder signals do not imply tighter constraints on $\vartheta$.
In particular, consider the (re)parametrization of $h$ from Sec.~\ref{sec:searches and sensitivity} ($h = a \tilde{h}$).
For any parameter besides $a$, all factors of $a$ cancel.
For $a$ itself, $I_{aa} \sim a^{-2}$ and $\mathrm{Var}(a) \propto a^2$, implying a constant fractional uncertainty in $a$.

Contrast this with the naive estimate that neglects calibration errors, where one expects $I_{ab} \sim |h|^2$.
The naive estimate suggests exceptionally well measured parameters in the limit of loud signals.
This cannot be the case in any physically realizable detector, as it is impossible to perfectly constrain the detector response.
Indeed, the smallest principle value of $\Gamma$ sets a lower limit on how well individual astrophysical parameters can be measured.\footnote{If there are prefect correlations between frequencies, as in Sec.~\ref{sec:big correlations}, then the smallest eigenvalue of $\Gamma$ is zero.}
I revisit this point in Sec.~\ref{sec:conclusion}

\section{Waveform Uncertainty and Deviations from General Relativity}
\label{sec:waveform uncertainty and deviations from general relativity}

Finally, as has been pointed out previously (e.g., Refs.~\cite{Lindblom:2008, Lindblom:2009, Purrer2019}), deviations in the waveform $h_\alpha \rightarrow h_\alpha(1 + \delta)$ produce analogous terms in the likelihood: $h = f_\alpha h_\alpha \rightarrow h(1+\delta)$.
If one can express the prior uncertainty on waveform errors or possible deviations from GR in terms of a GP, then the expressions derived in the context of calibration uncertainty, to a large part, hold unaltered.
The interpretation of $\gamma$ and $\Gamma$ changes, but the mechanics of how to constrain them and how they might affect the inference remain the same.
Just as one is able to analytically integrate out calibration uncertainties to obtain a marginal likelihood for the data, one can marginalize over uncertainty in the astrophysical signal to obtain the same.
In this sense, the waveform constitutes a probabilistic map between a system's properties (masses, spins, etc) and the expected signal from the true theory of gravity.
Of particular interest may be hierarchical inference for the distribution of waveform errors assuming these errors are independently and identically distributed draws from the same GP prior for each event.
This would be, in effect, a nonparametric extension to the parametrized hierarchical analysis proposed in Ref.~\cite{Isi:2019} and used extensively within Refs.~\cite{GWTC2_TGR, GWTC3_TGR}.

A similar approach was explored in Ref.~\cite{Edelman:2021} using cubic splines to parametrize deviations from GR, although they did not perform hierarchical inference and only analyzed events separately.
While splines are related to GPs~(see Chapter 6 in Ref.~\cite{Rasmussen:2006}), a direct formulation of the problem in terms of GPs will likely allow for more control over prior assumptions and improved interpretability of the resulting constraints.

Although I leave a thorough investigation of the implications of GP priors for waveform errors and deviations from GR to future work, I note a few things.
In this case, one expects perfect correlations in $\delta$ between IFOs (all detectors witness the same astrophysical signal) and the deviations to be drawn from the same distribution for all events (the signal of each event is determined by the same underlying theory of gravity).
However, one may need to allow for separate deviations for each polarization ($\delta_\alpha$ may be different for each $\alpha$), mixing between polarizations ($\delta_\alpha$ may depend on $h_\beta$), and these deviations may not be best expressed as fractional changes in the waveform ($\delta_\alpha$ may be nonzero even when $h_\alpha = 0 \ \forall \ \alpha$).

As a final note, one may consider GP priors for both calibration ($\delta_f$) and waveform ($\delta_h$) uncertainty simultaneously.
In this case, the projected strain in the detectors appears as $h = f_\alpha h_\alpha \rightarrow h(1+\delta_f)(1+\delta_h)$.
The likelihood will contain a term like $|h \delta_f \delta_h|^2$, which may stymie analytic marginalization over both $\delta_f$ and $\delta_h$, although one should still be able to analytically marginalize over one of them.
Again, I leave a full investigation to future work.

\section{Discussion}
\label{sec:conclusion}

\begin{table*}
    \caption{
        Estimates of the precision of individual parameters for a 1.4+1.4 $M_\odot$ binary when $l_\Gamma=0$ with projected design sensitivities of future IFOs.
        I show the uncertainty assuming no calibration error ($\sigma_\Gamma \rightarrow 0$) at a reference S/N of 10 as well as the lower limit set by nonzero calibration uncertainty.
        The lower limits scale linearly with the calibration uncertainty (and the square root of the frequency spacing), and I quote reference values for $\sigma_\Gamma \sqrt{\Delta f} = 0.01$.
        In addition to the uncertainty in the coalescence time ($t_c$), I approximate the corresponding uncertainty in the polar angle (localization precision) between the LIGO Hanford (LHO) and LLO detectors from triangulation ($\sigma_{\theta_{HL}} \approx \sqrt{2} (\sigma_{t_c} / 10\,\mathrm{ms})$)~\cite{Fairhurst:2010}.
    }
    \label{tab:sensitivity limits}
    {\renewcommand{\arraystretch}{1.8}
    \begin{tabular}{@{\extracolsep{0.75cm}} c c c c c c c}
        \hline
        \hline
        IFO & quantity & $\mathcal{M}\,[M_\odot]$ & $\eta$ & $\tilde\Lambda$ & $t_c\,[\mathrm{ms}]$ & $\theta_{HL}$ \\
        \hline
        aLIGO design~\cite{aLIGOpsd}
          & \multirow{2}{*}{$\left(\rho/10\right) \lim\limits_{\Gamma\rightarrow0} \sigma_i$}
            & \result{$1.06\cdot10^{-4}$} & \result{$3.17\cdot10^{-3}$} & \result{$2.61\cdot10^{2}$} & \result{$3.14\cdot10^{-1}$} & \result{$2.5^\circ$} \\
        Cosmic Explorer~\cite{CEpsd}
          &
            & \result{$7.46\cdot10^{-5}$} & \result{$2.24\cdot10^{-3}$} & \result{$5.09\cdot10^{+2}$} & \result{$5.66\cdot10^{-1}$} & \result{$4.6^\circ$} \\
        \hline
        $f \in [5,1500]\,\mathrm{Hz}$
          & $\left( 0.01 / \sigma_\Gamma \sqrt{\Delta f} \right) \lim\limits_{\rho\rightarrow\infty} \sigma_i$
            & \result{$1.44\cdot10^{-6}$} & \result{$4.31\cdot10^{-5}$} & \result{$4.27\cdot10^{-1}$} & \result{$1.03\cdot10^{-3}$} & \result{$30.0^{\prime\prime}$} \\
        \hline
    \end{tabular}
    }
\end{table*}

GW astrophysics has rapidly grown over the last few years with an ever increasing number of confidently detected compact binary coalescences~\cite{GWTC1, GWTC2, GWTC2d1, GWTC3} and steadily improved upper limits on the rates of other sources (see, e.g.,~\cite{O3CWknown, O3CWisolated}).
This has been enabled in no small part by the remarkable understanding of the current interferometers' responses to astrophysical strain.
Indeed, in addition to providing accurate calibration for each interferometer in the network, analysts are able to carefully quantify the precision of their calibration estimates through GPs, and the incorporation of this information within astrophysical inference has become commonplace within the literature.

This manuscript exploits the fact that the dominant calibration uncertainties are described by GPs to analytically marginalize over them.
Although several authors have already demonstrated numeric marginalization techniques, this analytic approach offers several advantages.

First, it may be computationally faster in practice.
That is, it may be faster to compute a single inner product than to directly Monte Carlo sample over the GP calibration prior.
The limiting step will likely be the inversion of $\Gamma$ and $A^{-1} = H^\mathrm{T} C^{-1} H + \Gamma^{-1}$ required to compute $B^{-1}$ for each signal model. This step may scale as $O(N_\mathrm{frq}^3)$.
Monte Carlo integration could be faster if very few Monte Carlo samples are needed for the integral to converge.
This will likely be the case only when the calibration prior is already so precise that it is not changed significantly by the addition of astrophysical data (the current situation for all signals detected to date).
That is, direct numeric integration may only be faster when the calibration uncertainties are too small to make much of a difference anyway.
At the same time, one may also be able to approximate $A \approx \Gamma$ in this limit, removing the need to invert $\Gamma$ and $A$ and thereby reducing the computational cost from $O(N_\mathrm{frq}^3)$ to $O(N_\mathrm{frq}^2)$.

What's more, analytic expressions allow for greater interpretability of the resulting conditioned and marginal distributions.
For example, I am able to easily take several limits and explore the impact of large/small calibration uncertainties and correlations.
I also show how to trivially extend the formalism to include networks of multiple detectors and multiple astrophysical signals, including possible correlations between the calibration errors witnessed in each detector/event pair.
Equivalent investigations via numeric integration are possible, but may be more difficult to interpret and more costly to perform.

Specifically, I find that calibration uncertainties always reduce the measurability of astrophysical parameters.
This can be seen in several ways.
I explore the impact of calibration uncertainties on search sensitivity through the behavior of the standard matched-filter S/N, finding larger variance in the presence of calibration uncertainty and, therefore, reduced search sensitivity.
Monte Carlo simulations of search sensitivity~\cite{Tiwari:2018, Farr:2019, Essick:2021oct} may be able to exploit this analytic marginalization to better account for calibration uncertainties without recourse to more direct numerical treatments.
However, perhaps the most striking example can be seen through the change in the Fisher information matrix elements.

The loss of information is a projection of the signal onto the calibration uncertainty's prior covariance matrix.
That is, the additional uncertainty in astrophysical parameters is directly related to whether the types of changes in the waveform induced by changing those parameters are common calibration errors.
Additionally, it is not necessary to constrain the calibration uncertainty to be small at all frequencies for accurate inferences of individual signals.
Instead, one need only constrain particular combinations of frequencies in order to minimize the information lost.
This also suggests an algorithm to determine the frequencies at which one should measure the calibration more precisely in order to best reduce the uncertainty in astrophysical parameters.
As a rule of thumb, these are (unsurprisingly) the frequencies at which the current calibration uncertainty most hurts the inference.

When calibration errors are large (and correlations between frequencies are not perfect), the uncertainty in astrophysical parameters becomes independent of the signal amplitude.
The parameters of louder signals are not better constrained than the parameters of quieter signals.
This sets a limit on the precision to which the exceptionally loud astrophysical signals can be measured.
The relevant comparison is whether the S/N is greater or smaller than the inverse of the calibration error's standard deviation ($\rho \lessgtr \sigma_\Gamma^{-1}$).
Including an approximate proportionality constant, I find that $1.4+1.4\,M_\odot$ binaries with $\rho \gtrsim 10^3$ require calibration uncertainties $\lesssim 3\%$.

\subsection{Estimates of calibration requirements for current and proposed detectors}
\label{sec:calibration requirements}

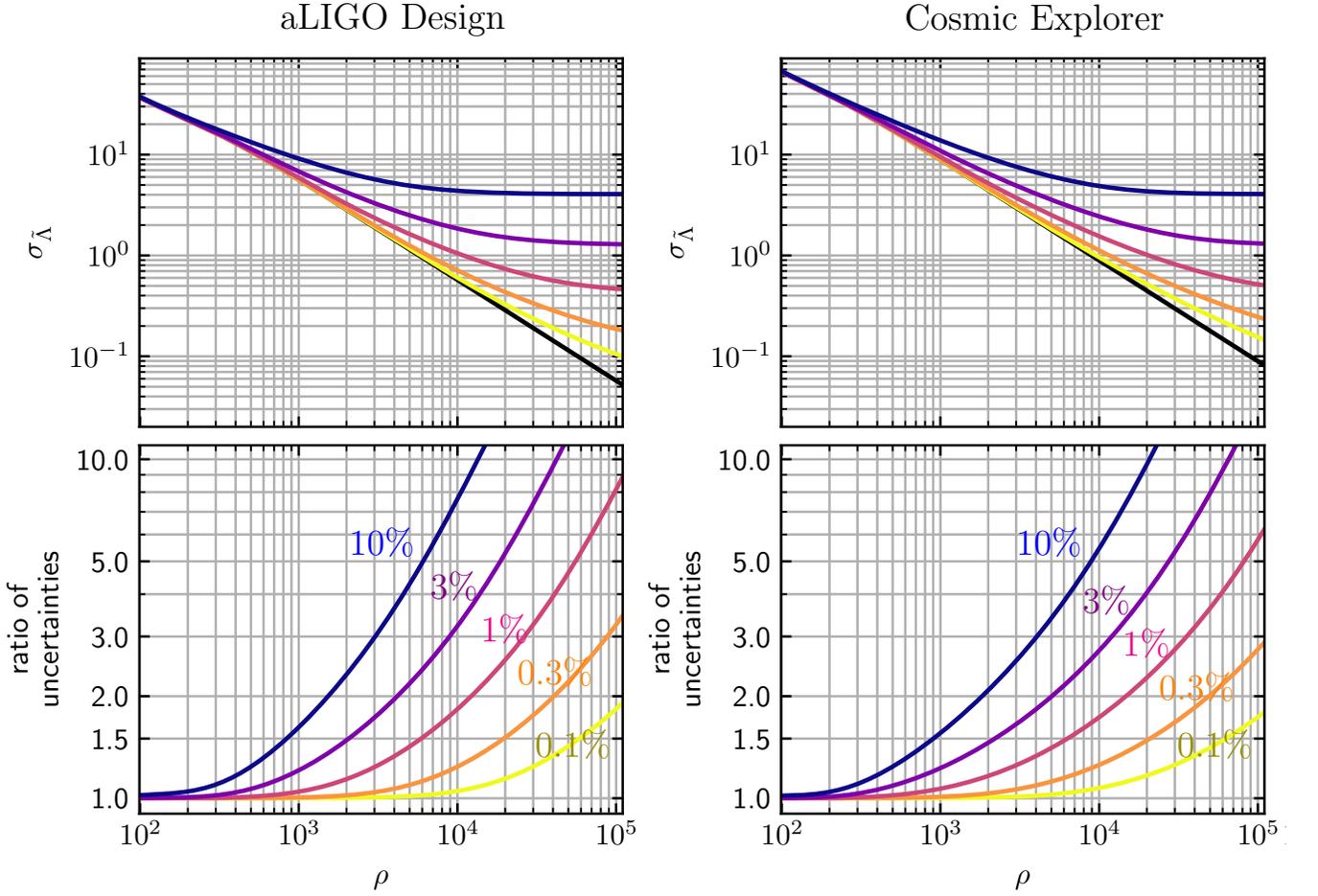
\begin{figure*}
    \begin{tikzpicture}
        \node (aligo) {\includegraphics[width=0.5\textwidth, clip=True, trim=0.0cm 0.0cm 7.6cm 0.0cm]{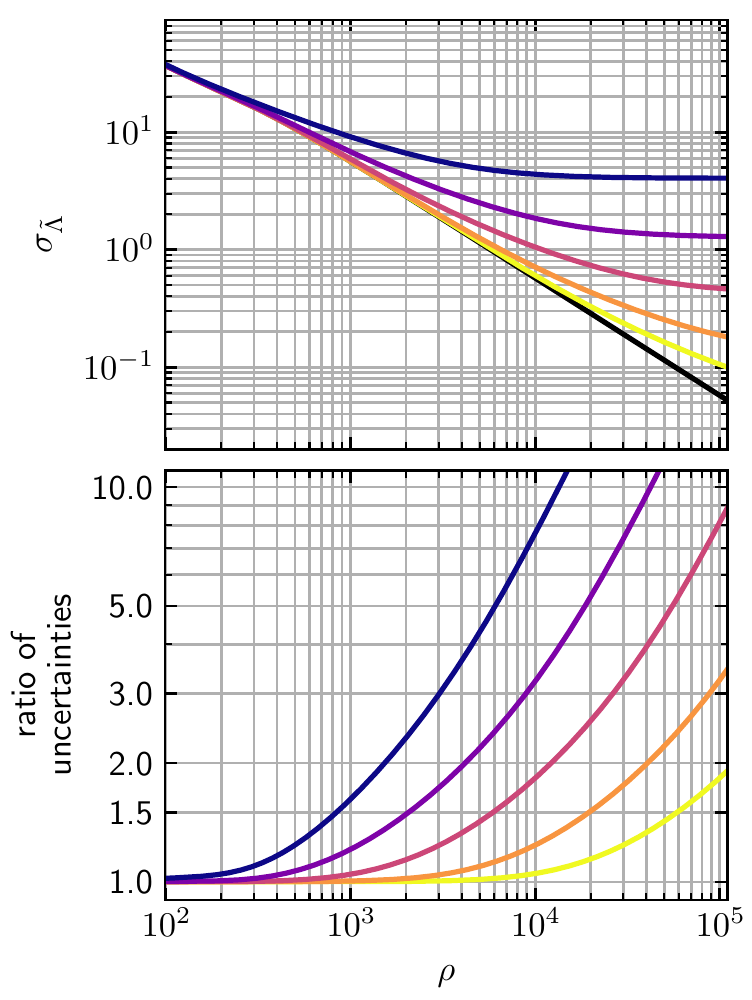}} ;
        \node (aligo-label) [above of=aligo, yshift=+5.25cm, xshift=+1.0cm] {\Large{aLIGO Design}} ;
        \node (aligo-10) [below of=aligo, xshift=+0.85cm] {\textcolor{blue}{\Large{10\%}}} ;
        \node (aligo-3) [below of=aligo, xshift=+1.85cm, yshift=-0.6cm] {\textcolor{violet}{\Large{3\%}}} ;
        \node (aligo-1) [below of=aligo, xshift=+2.55cm, yshift=-1.2cm] {\textcolor{magenta}{\Large{1\%}}} ;
        \node (aligo-03) [below of=aligo, xshift=+3.25cm, yshift=-1.8cm] {\textcolor{orange}{\Large{0.3\%}}} ;
        \node (aligo-01) [below of=aligo, xshift=+3.50cm, yshift=-2.8cm] {\textcolor{olive}{\Large{0.1\%}}} ;

        \node (ce) [right of=aligo, xshift=+7.90cm] {\includegraphics[width=0.5\textwidth, clip=True, trim=0.0cm 0.0cm 7.6cm 0.0cm]{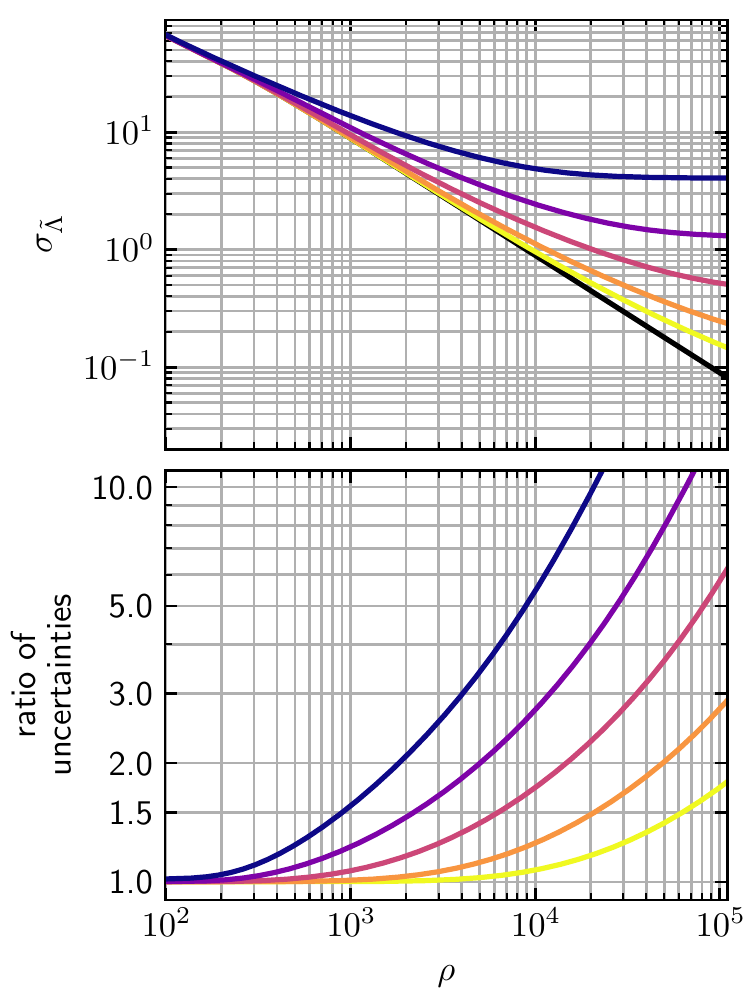}} ;
        \node (ce-label) [above of=ce, yshift=+5.25cm, xshift=+1.0cm] {\Large{Cosmic Explorer}} ;
        \node (ce-10) [below of=ce, xshift=+1.20cm] {\textcolor{blue}{\Large{10\%}}} ;
        \node (ce-3) [below of=ce, xshift=+2.00cm, yshift=-0.8cm] {\textcolor{violet}{\Large{3\%}}} ;
        \node (ce-1) [below of=ce, xshift=+2.55cm, yshift=-1.4cm] {\textcolor{magenta}{\Large{1\%}}} ;
        \node (ce-03) [below of=ce, xshift=+3.25cm, yshift=-2.0cm] {\textcolor{orange}{\Large{0.3\%}}} ;
        \node (ce-01) [below of=ce, xshift=+3.50cm, yshift=-2.8cm] {\textcolor{olive}{\Large{0.1\%}}} ;
    \end{tikzpicture}
    \caption{
        (\emph{top}) Estimates of the posterior uncertainty in $\tilde\Lambda$ and (\emph{bottom}) ratios of uncertainty with and without calibration errors for a 1.4+1.4 $M_\odot$ system as a function of S/N with (\emph{left}) aLIGO design sensitivity~\cite{aLIGOpsd} and (\emph{right}) a reference Cosmic Explorer design~\cite{CEpsd}.
        I show the limit $\l_\Gamma=0$ with (\emph{dark to light}) $\sigma_\Gamma \sqrt{\Delta f}=10\%$, 3\%, 1\%, 0.3\%, and 0.1\%.
        Due to Cosmic Explorer's higher sensitivity at lower frequencies, observations with Cosmic Explorer constrain $\tilde\Lambda$ less than an equivalent observation (same S/N) with aLIGO.
        However, the distribution of detected sources expected for Cosmic Explorer extends to higher S/N than for aLIGO.
        I find that, for both detectors, one requires $\rho \gtrsim 10^4$ to obtain $\sigma_{\tilde\Lambda} \lesssim 1$ with reasonable calibration uncertainty ($\sigma_\Gamma \sqrt{\Delta f} \gtrsim 1\%$), and, at this S/N, the contribution to $\sigma_{\tilde\Lambda}$ from calibration uncertainty is at least as large as the contribution from stationary Gaussian noise (ratio of uncertainty $\gtrsim 2$).
    }
    \label{fig:tides}
\end{figure*}

Although a full description of the impact of all possible calibration priors on astrophysical inference with all possible detectors is beyond the scope of this study, I nonetheless attempt to provide some quantitative estimates of a few limiting cases for a few detectors.
Specifically, I consider independent calibration uncertainty for the real and imaginary parts of the calibration error (uncorrelated amplitude and phase errors), each of which is described by a zero-mean squared-exponential GP with the same hyperparameters:
\begin{align}
    \mathrm{Cov}(\mathcal{R}\{\delta(f)\}, \mathcal{R}\{\delta(f^\prime)\})
        & = \mathrm{Cov}(\mathcal{I}\{\delta(f)\}, \mathcal{I}\{\delta(f^\prime)\}) \nonumber \\
        & = \sigma_\Gamma^2 \exp\left( -\frac{(f - f^\prime)^2}{l_\Gamma^2}\right) \\
    \mathrm{Cov}(\mathcal{R}\{\delta(f)\}, \mathcal{I}\{\delta(f^\prime)\})
        & = 0 \ \forall \ f, f^\prime
\end{align}
With this definition of calibration uncertainty, I then estimate the expected marginal uncertainty in a few astrophysical parameters by computing the Fisher information matrix.
As shown in the Appendix, the limit $l_\Gamma \rightarrow 0$ corresponds to a local maximum in the uncertainty in astrophysical parameters.
I therefore focus on this limit as a worst case scenario given a marginal uncertainty at each frequency (fixed $\sigma_\Gamma$).
Nonzero (but finite) $l_\Gamma$ may reduce astrophysical uncertainty by a factor of a few.

Furthermore, I consider a \result{$1.4+1.4\,M_\odot$ non-spinning BNS system with each component's tidal deformability $\Lambda = 500$}.
I approximate the frequency data as a regularly spaced array from \result{$5\,\mathrm{Hz}$} to \result{$1500\,\mathrm{Hz}$}, with a frequency spacing of \result{$\Delta f = 1\,\mathrm{Hz}$}.
My waveform model assumes the astrophysical parameters only affect the signal's phase, and I describe the phase in the frequency domain with a post-Newtonian (PN) sum.
That is, I model the waveform following Ref.~\cite{Flanagan:2008}
\begin{equation}
    h = a f^{-7/6} e^{i\psi(f)}
\end{equation}
where
\begin{align}
    \psi(f) 
        & = 2\pi f t_c
        - \frac{117}{256} \tilde\Lambda \frac{v^5}{\eta} \nonumber \\
        & \quad + \frac{3}{128 \eta v^5} \left[
            1 
            + \frac{20}{9} \left( \frac{743}{336} + \frac{11}{4}\eta \right) v^2
            - 4(4\pi - \beta) v^3 \right. \nonumber \\
        & \quad \quad \quad \quad \left.
            + 10 \left( \frac{3058673}{1016064} + \frac{5429}{1008}\eta + \frac{617}{144} \eta^2 - \sigma \right) v^4 \right] \nonumber \\
\end{align}
and
\begin{gather}
    M_\mathrm{tot} = m_1 + m_2 \\
    \eta = \frac{m_1 m_2}{M_\mathrm{tot}^2} \\
    \mathcal{M} = \eta^{3/5} M_\mathrm{tot} \\
    v = (\pi M_\mathrm{tot} f)^{1/3} \\
    \tilde\Lambda = \frac{16}{13}\left( \frac{(m_1 + 12m_2)m_1^4 \Lambda_1}{M_\mathrm{tot}^5} + 1 \leftrightarrow 2 \right)
\end{gather}
$\beta$ and $\sigma$ are spin parameters.
Although more accurate waveforms are available, the simplicity of this model's phasing (and derivatives thereof) make it well-suited to my particular estimation task.
I estimate the uncertainty for $\mathcal{M}$, $\eta$, $\beta$, $\sigma$, $\tilde\Lambda$, and $t_c$ via the inverse of the Fisher information matrix.
In the limit $l_\Gamma=0$ and assuming the parameters only affect the waveform phase, the Fisher matrix elements are
\begin{equation} \label{eq:fisher l=0}
    I_{ab}^{(l_\Gamma=0)} = 4\int df\, \left(\frac{\partial_a \psi \partial_b \psi |h|^2}{s} \right) \left( 1 + \frac{4\Delta f \sigma_\Gamma^2 |h|^2}{s} \right)^{-1}
\end{equation}

Additionally, I apply Gaussian priors for the two spin parameters $\beta$ and $\sigma$ with standard deviations of 0.05 (small spins) and for the symmetric mass ratio with a standard deviation of 0.25 (physical limit).
These priors are centered on the injected values.
No priors were applied for the $\mathcal{M}$, $\tilde\Lambda$, or $t_c$.
This setup approximately reproduces the uncertainty reported for GW170817~\cite{GW170817Properties} at the corresponding S/N.

Given the various approximations within my calculation, in addition to the standard caveats about the reliability of the Fisher matrix~\cite{Vallisneri:2008}, I must stress that these calibration requirements should be interpreted as informed estimates rather than rigorous bounds.
Nonetheless, I present an analysis of expected uncertainties assuming aLIGO design sensitivity~\cite{LIGO, aLIGOpsd} in Sec.~\ref{sec:aligo design} and of a nominal Cosmic Explorer design~\cite{NextGeneration, CEpsd, CosmicExplorer, Reitze:2019} in Sec.~\ref{sec:cosmic explorer}.
The Appendix describes these results in more detail.

\subsubsection{Forecasts for aLIGO design sensitivity}
\label{sec:aligo design}

Table~\ref{tab:sensitivity limits} presents several estimates of the the expected uncertainty for a $1.4+1.4\,M_\odot$ binary in two limits of calibration uncertainty.
First, it shows the limit of zero calibration uncertainty ($\sigma_\Gamma=0$), in which case the uncertainty scales inversely with S/N.
At a nominal S/N of 10 (approximate detection threshold), I find $\sigma_\mathcal{M} \sim O(10^{-4}\,M_\odot)$ and $\sigma_{\tilde\Lambda} \sim \mathrm{few} \cdot 10^2$.

Table~\ref{tab:sensitivity limits} also shows the lower limit in the uncertainty in astrophysical parameters set by the calibration uncertainty when $l_\Gamma = 0$.
This scales with the calibration uncertainty ($\sigma_\Gamma \sqrt{\Delta f}$, see Appendix), and I quote values at nominal 1\% calibration uncertainty.
A back-of-the-envelope estimate for the S/N at which calibration uncertainty becomes important can be obtained by equating the two cases.
For example, this yields
\begin{equation} \label{eq:aligo snr thresh}
    \rho = 6110 \left(\frac{0.01}{\sigma_\Gamma\sqrt{\Delta f}}\right)
\end{equation}
for $\tilde\Lambda$ with aLIGO.

As a general rule, the posterior uncertainty in $\mathcal{M}$ and $\eta$ become limited by calibration uncertainty at
lower S/N than $\tilde\Lambda$.
The equivalent of Eq.~\ref{eq:aligo snr thresh} for $\mathcal{M}$ has a coefficient of 736 (nearly an order of magnitude smaller).
However, the mass parameters tend to be so well measured anyway that the limits imposed by calibration uncertainty are
 too small to be of practical importance.
I therefore focus on the leading order tidal parameter ($\tilde\Lambda$) when quoting approximate calibration requirements.
For example, I expect to never measure $\tilde\Lambda$ to better than $\pm 0.5$ when $\sigma_\Gamma\sqrt{\Delta f} = 1\%$.
While this is still exceptional precision ($\sim 0.1\%$ relative uncertainty~\cite{Legred:2021}), it is only reached in the limit of extremely loud signals: $\rho > 10^5$, implying a Galactic merger $O(10\,\mathrm{kpc})$ away.

Figure~\ref{fig:tides} builds upon this by showing the scaling between $\sigma_{\tilde\Lambda}$ and S/N for various $\sigma_\Gamma$.
From these curves, one can read off the S/N required to reach a particular level of precision if one knows the calibration uncertainty.
Alternatively, if one knows the loudest S/N that is likely to be observed, they can read off the level of calibration uncertainty that can be tolerated before it dominates over uncertainty from the stationary Gaussian noise.
The lower panels in Fig.~\ref{fig:tides} show this explicitly as the ratio of $\sigma_{\tilde\Lambda}$ marginalized over calibration uncertainty to $\sigma_{\tilde\Lambda}$ assuming perfect calibration.
As a rule of thumb, one might require the calibration uncertainty to add no more than 50\% more relative uncertainty compared to stationary Gaussian noise alone.
With this criterion, we see that $\sigma_\Gamma \sqrt{\Delta f} \lesssim 10\%$ suffices for $\tilde\Lambda$ when $\rho \leq 10^3$.
Given the expected S/N distribution for aLIGO ($p(\rho) \propto \rho^{-4}$) and an approximate detection threshold of 10, this implies that $\sigma_\Gamma \sqrt{\Delta f} \lesssim 10\%$ will produce less than 50\% additional uncertainty for all but 1 in $10^6$ detected signals.

\subsubsection{Forecasts for Cosmic Explorer}
\label{sec:cosmic explorer}

Table~\ref{tab:sensitivity limits} and Fig.~\ref{fig:tides} show that the behavior (as a function of S/N) is remarkably similar for both aLIGO design sensitivity and Cosmic Explorer.
The main difference is that Cosmic Explorer is has more sensitivity at lower frequencies and therefore measures $\mathcal{M}$ to higher precision and $\tilde\Lambda$ to lower precision compared to similar S/N signals in aLIGO.
Generally, I find comparable rules of thumb apply regarding calibration requirements.
The expected S/N distribution for $1.4+1.4\,M_\odot$ binaries detected with Cosmic Explorer when signals are distributed uniformly in comoving volume (and source-frame time) is similar to the distribution detected with aLIGO.
As such, the calibration will affect approximately the same fraction of low-mass mergers in both aLIGO and Cosmic explorer.

It is worth noting, though, that calibration requirements for BBHs may be more stringent.
Although my waveform model is unsuitable for precise BBH predictions (I truncate my waveform sharply and ignore merger and ringdown~\cite{Mandel:2014}), one might expect stricter calibration requirements for BBH than for BNS at the same S/N because the BBH signal bandwidth is smaller.
That is, $|h|$ be larger at each frequency for BBH, and therefore the relative importance of $\Gamma \neq 0$ will be larger.
However, as much of a BNS's S/N is gathered at low frequencies, just like a BBH, it is unlikely that this would produce an effect larger than a factor of a few on calibration requirements.

Of perhaps greater astrophysical interest is the fact that the distribution S/N for BBH detected with Cosmic Explorer will not be peaked near the detection threshold~\cite{Vitale:2016}.
This effect is also present for BNS, but it is much more pronounced for BBH.
One may reasonably expect a significant fraction of detected BBH to have $\rho \gtrsim 10^2$.
However, the distribution of the loudest sources will still be limited by the amount of comoving volume at low redshift (assuming a constant merger rate in the local universe).
Thus, the tail of the distribution of observed S/N will still fall off as $\rho^{-4}$ (Euclidean universe).
If calibration uncertainties are small enough to not matter until $\rho \gtrsim 10^3$, then they will not matter for the vast majority of detectable binaries even if the distribution of detected S/N peaks near 100.

\subsection{Next Steps}
\label{sec:next steps}

Given the readily apparent benefits of this analytic approach to the impact of calibration uncertainties within GW inference, it may be of interest to extend this work in the future.
Specifically, it may be of interest to more closely examine the implications of Eq.~\ref{eq:delta | d} to understand what types of astrophysical signals may be able to most improve our understanding of calibration uncertainties.
It may also be of interest to further explore ways to represent parametrized models of the detector response in terms of GPs (e.g., through correlations between frequencies in $\Gamma$).
For example, Bayesian linear regression is a special case of a GP with a covariance kernel $K(x_i, x_j) \propto x_i x_j$.
That is, covariance kernels of this kind generate polynomial functions of $x$ without the need to specify these as part of a parametrized mean.
If this is possible, one may still be able to exploit the analytic marginalization derived in this manuscript that relies on expressing the calibration uncertainty as a GP even when the systematic errors in the modeled detector response shrink below the statistical precision to which the modeled response is measured.

Finally, I again note that the additional terms within the likelihood that are created by calibration errors can also be introduced by waveform errors or deviations from GR.
If one is able to express waveform uncertainty (or uncertainty in the true theory of gravity) in terms of a GP, analogous analytic treatments will be possible.
Indeed, with a proper choice of hyperprior for our uncertainty in the true theory of gravity, one may be able to perform a completely analytic hierarchical inference for the distribution of deviations from GR.
I will explore these possibilities further in future work.


\acknowledgments

I am very grateful for the useful discussions with Ethan Payne and the broader parameter estimation and calibration groups within the LIGO Scientific Collaboration.
I also thank the Canadian Institute for Advanced Research (CIFAR) for support.
Research at Perimeter Institute is supported in part by the Government of Canada through the Department of Innovation, Science and Economic Development Canada and by the Province of Ontario through the Ministry of Colleges and Universities.
This material is based upon work supported by NSF's LIGO Laboratory which is a major facility fully funded by the National Science Foundation.


\bibliography{refs}


\newpage
\onecolumngrid


\appendix


\section{Marginal Uncertainty when $\Gamma$ is not diagonal}
\label{sec:nonzero l}

\begin{figure*}
    \begin{minipage}{0.48\textwidth}
        \begin{center}
            \Large{stationary over $f$} \\
            \includegraphics[width=1.0\textwidth]{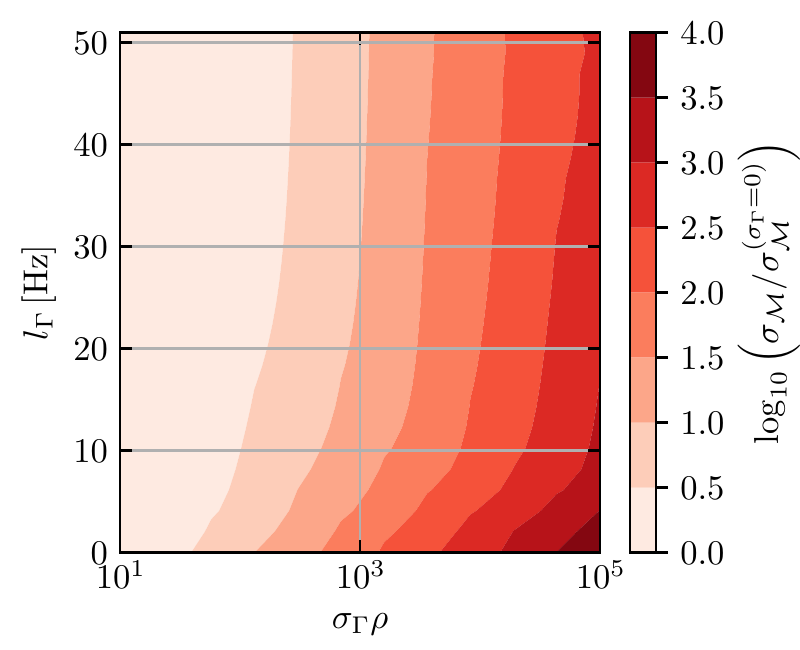}
        \end{center}
    \end{minipage}
    \hspace{0.005\textwidth}
    \begin{minipage}{0.48\textwidth}
        \begin{center}
            \Large{stationary over $\log_{10}(f)$} \\
            \includegraphics[width=1.0\textwidth]{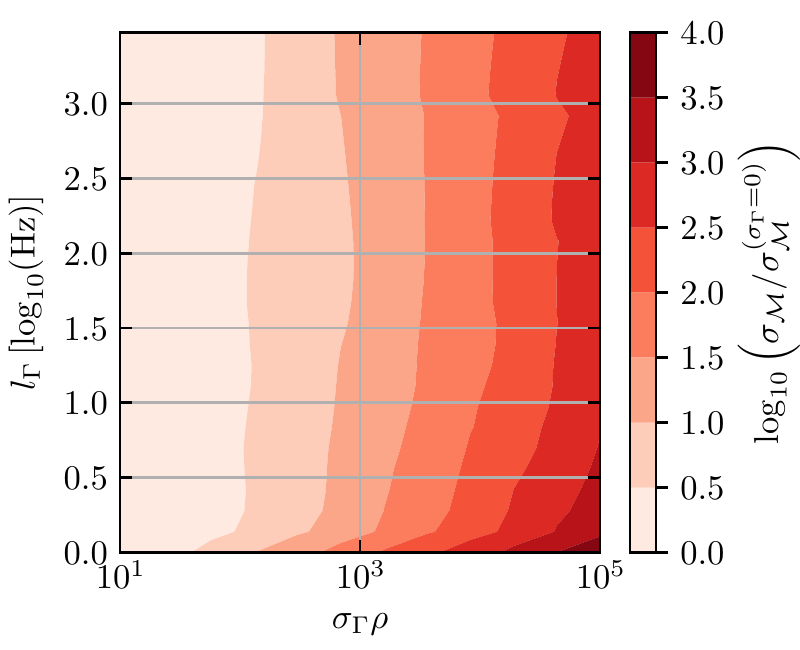}
        \end{center}
    \end{minipage}
    \caption{
        Estimates of $\sigma_\mathcal{M}$ for Cosmic Explorer~\cite{CEpsd} as a function of the calibration uncertainty when I assume correlations between frequencies in the calibration GP are stationary over (\emph{left}) the linear frequency difference and (\emph{right}) the logarithmic frequency difference (ratio of frequencies).
        I estimate these uncertainties by the inverse of the Fisher information matrix neglecting the trace term in Eq.~\ref{eq:multivariate Gaussian fisher information} (exact only when $\Gamma$ is diagonal).
        These are scaled by the naive estimate that assumes perfect calibration ($\sigma_\Gamma=0$) to remove the trivial scaling with S/N.
        For computational expediency, I only include frequencies up to 500 Hz (as opposed to 1500 Hz used elsewhere).
        Both types of correlations produce similar behavior, and similar results are found for all astrophysical parameters (modulo the impact of priors).
    }
    \label{fig:sigma-l grids}
\end{figure*}

In this appendix, I briefly investigate the behavior of the calibration-marginalized Fisher information matrix assuming squared-exponential GP priors for the calibration uncertainty with non-vanishing correlations between frequencies ($l_\Gamma \neq 0$).
Note that the Fisher information matrix elements can be scaled so that they only depend on two parameters
\begin{equation}
    \frac{1}{\rho^2} I_{ab} = \partial_a \tilde{h}^\mathrm{T} \tilde{B}^{-1} \partial_b \tilde{h} - \frac{1}{2}\mathrm{tr}\left[ \partial_a \tilde{B} \partial_b \tilde{B}^{-1} \right]
\end{equation}
where
\begin{gather}
    h = a \tilde{h} \\
    H = a\tilde{H} \\
    \tilde{h}^\mathrm{T} C^{-1} \tilde{h} = 1 \\
    \tilde{B}^{-1} = C^{-1} - C^{-1} \tilde{H} \tilde{A} \tilde{H}^\mathrm{T} C^{-1} \\
    \tilde{A}^{-1} = \tilde{H}^\mathrm{T} C^{-1} \tilde{H} + x \tilde{\Gamma}^{-1} \\
    \Gamma = \sigma_\Gamma^2 \tilde{\Gamma}
\end{gather}
and
\begin{equation}
    x \equiv \frac{1}{\sigma_\Gamma^2 \rho^2}
\end{equation}
Again, one sees that the comparison that determines whether calibration uncertainty is important within the astrophysical inference (whether $x$ is large or small) is between $\sigma_\Gamma$ and $\rho$.
This appendix is dedicated to illuminating $l_\Gamma$'s role as well.
Fig.~\ref{fig:sigma-l grids} shows the main results, which are
\begin{itemize}
    \item $l_\Gamma=0$ (diagonal $\Gamma$) is a local maximum in uncertainty,
    \item finite $l_\Gamma$ can reduce the uncertainty as much as a factor of $\sim 10$, and
    \item perfect correlations between frequencies (infinite $l_\Gamma$) no longer impose an absolute minimum on the uncertainty of astrophysical parameters.
\end{itemize}

In Fig.~\ref{fig:sigma-l grids} and throughout this appendix, I approximate $I_{ab} \approx \partial_a h^\mathrm{T} B^{-1} \partial_b h$ and neglect the trace term in Eq.~\ref{eq:multivariate Gaussian fisher information}.
This is only exact when $\Gamma$ is diagonal and $\mathrm{Var}\{\mathcal{R}\{\delta(f_i)\}\} = \mathrm{Var}\{\mathcal{I}\{\delta(f_i)\}\} \ \forall \ f_i$, in which case $B$ depends only on $|h|^2$ and therefore $\partial_a B = 0$ whenever the astrophysical parameters only affect the signal's phase.
The trace term is always negligible when $l_\Gamma$ is small, but it is not know whether it is important for larger $l_\Gamma$.
What is known, however, is that it is very difficult to accurately compute the trace term for large $l_\Gamma$ because $\Gamma$ becomes ill-conditioned (the ratio of the largest eigenvalue to the smallest eigenvalue is large) and therefore difficult to invert.
Even more so than the results in the main text, then, my conclusions about the behavior at large $l_\Gamma$ should be taken with a healthy grain of salt.

First, let's begin with the observation that diagonal $\Gamma$ correspond to local maxima in the uncertainty within the astrophysical inference.
Note that
\begin{equation}
    \frac{d}{dl_\Gamma} I_{ab} = \partial_a h^\mathrm{T} \frac{d B^{-1}}{dl_\Gamma} \partial_b h + \frac{1}{2}\mathrm{tr} \left[\frac{dB}{dl_\Gamma} \partial_a B^{-1} B \partial_b B^{-1} + B \frac{d(\partial_a B^{-1})}{dl_\Gamma} B \partial_b B^{-1} + B \partial_a B^{-1} \frac{dB}{dl_\Gamma} \partial_b B^{-1} + B \partial_a B^{-1} B \frac{d(\partial_b B^{-1})}{dl_\Gamma} \right]
\end{equation}
and 
\begin{equation}
    \left. \frac{d}{dl_\Gamma} I_{ab} \right|_{l_\Gamma=0} = \partial_a h^\mathrm{T} \left(\left.\frac{dB^{-1}}{dl_\Gamma}\right|_{l_\Gamma=0}\right) \partial_b h
\end{equation}
because $\partial_a B^{-1}|_{l_\Gamma=0} = 0$.
With further manipulation, one obtains
\begin{equation}\label{eq:this one}
    \frac{dB^{-1}}{dl_\Gamma} = - C^{-1} H A \Gamma^{-1} \left(\frac{d\Gamma}{dl_\Gamma}\right) \Gamma^{-1} A H^\mathrm{T} C^{-1}
\end{equation}
and, for my squared-exponential GPs
\begin{equation}
    \frac{d\Gamma_{ij}}{dl_\Gamma} = \left(\frac{2(x_i-x_j)^2}{l_\Gamma^3}\right) \Gamma_{ij} \geq 0 \ \forall \ i,j
\end{equation}
What's more, L'H\^{o}pital's rule shows that
\begin{equation}
    \lim\limits_{l_\Gamma \rightarrow 0^+} \frac{d\Gamma_{ij}}{dl_\Gamma} = 0 \ \forall \ i,j
\end{equation}
and therefore $(dI_{ab}/dl_\Gamma)|_{l_\Gamma = 0} = 0$.\footnote{Similar considerations show that $(d(\partial_a B^{-1})/dl_\Gamma)|_{l_\Gamma=0} = 0$.}

This makes sense, as diagonal $\Gamma$ maximize the entropy (provide the least information) within the calibration prior for each $\sigma_\Gamma$.
One expects the smallest amount of prior calibration information to correspond to a maximum in the uncertainty in the astrophysical inference.
As the correlations between frequencies increase, the types of calibration error allowed by the prior are more restricted, and this, almost certainly, reduces the uncertainty in the astrophysical parameters.

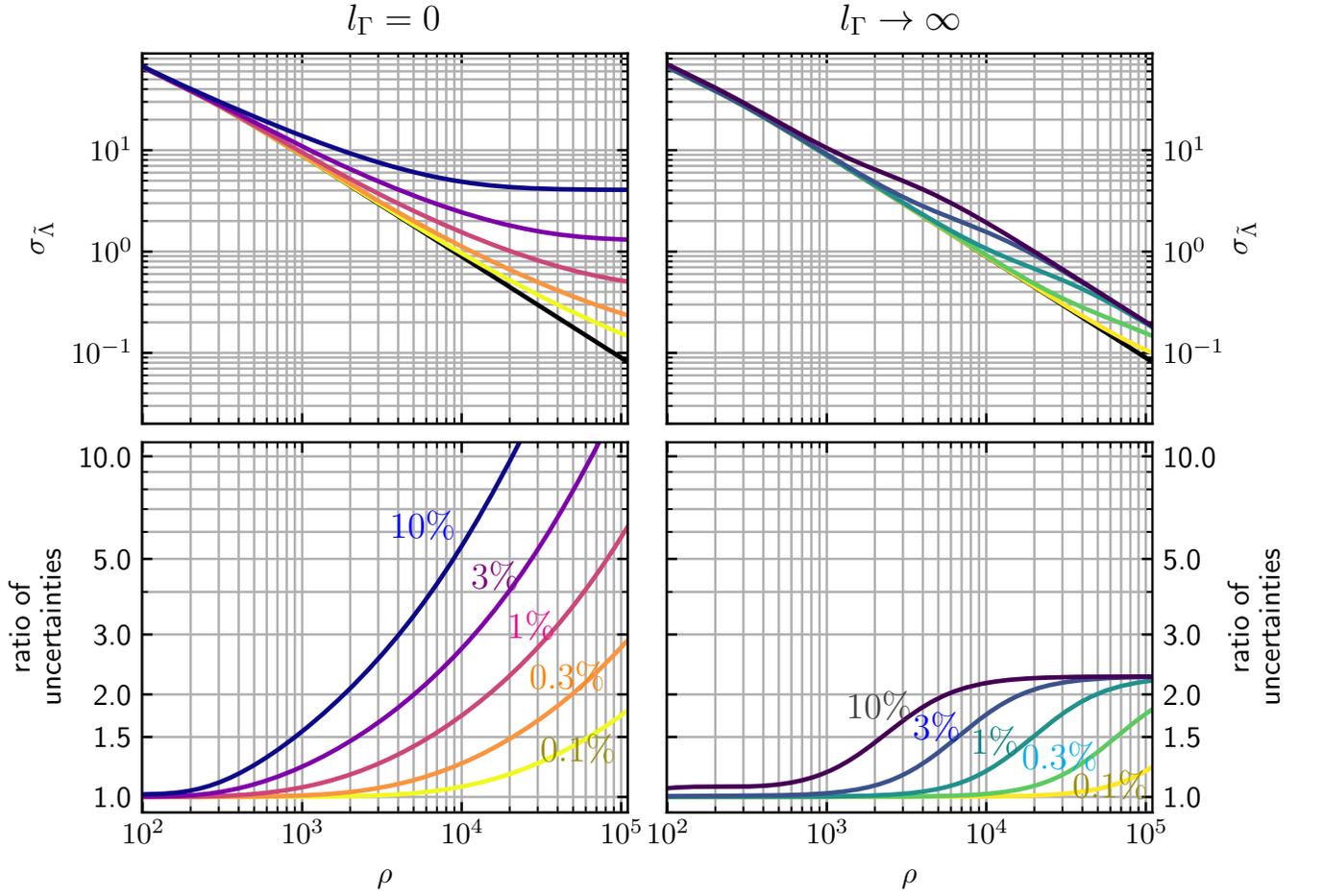
\begin{figure}
    \begin{tikzpicture}
        \node (panel) {\includegraphics[width=1.0\textwidth]{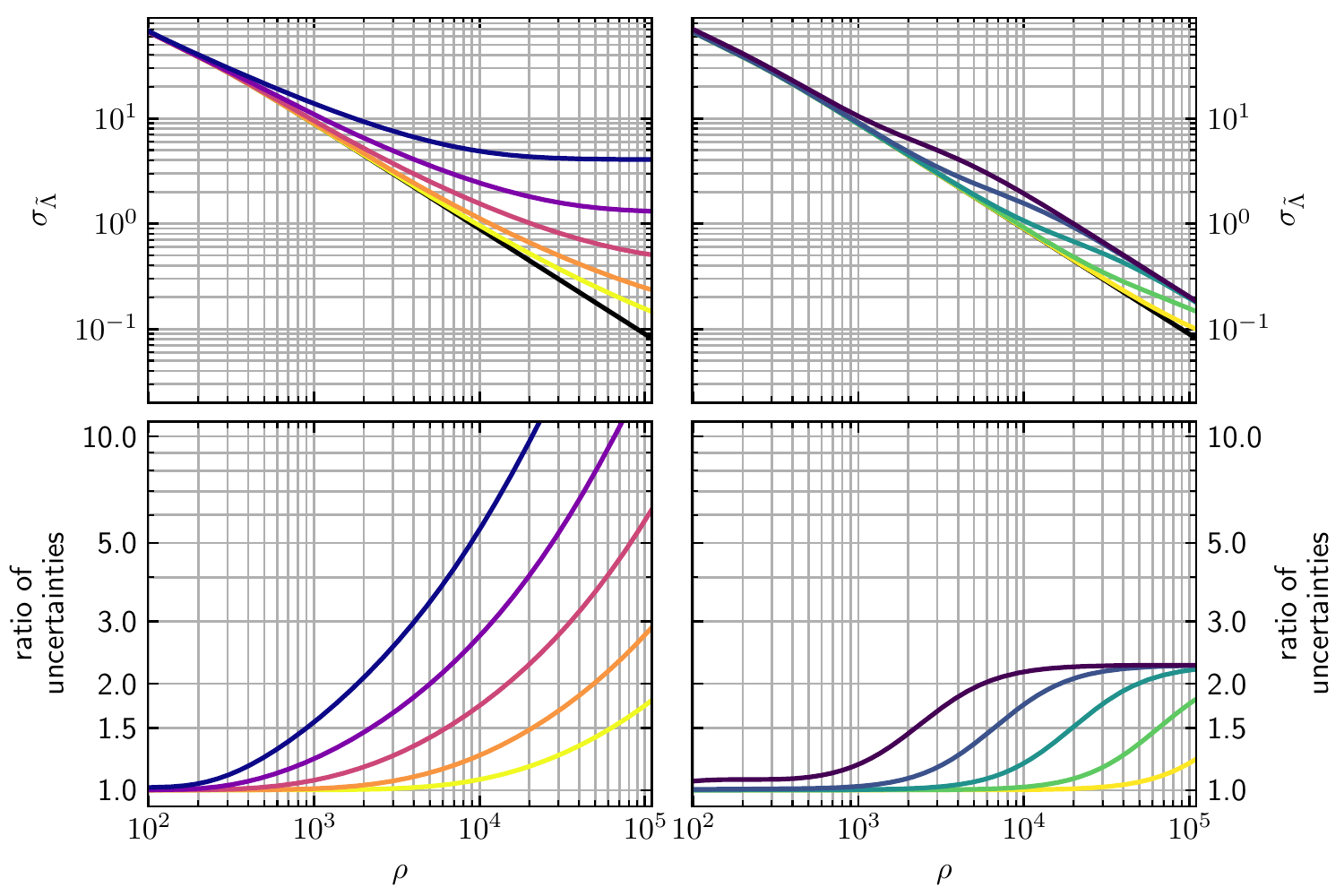}} ;

        \node (l0) [above left of=panel, yshift=+5.5cm, xshift=-2.8cm] {\Large{$l_\Gamma = 0$}} ;
        \node (l0-10) [below of=l0, xshift=+0.40cm, yshift=-6.0cm] {\textcolor{blue}{\Large{10\%}}} ;
        \node (l0-3) [below of=l0-10, xshift=+1.00cm, yshift=+0.3cm] {\textcolor{violet}{\Large{3\%}}} ;
        \node (l0-1) [below of=l0-3, xshift=+0.50cm, yshift=+0.3cm] {\textcolor{magenta}{\Large{1\%}}} ;
        \node (l0-03) [below of=l0-1, xshift=+0.50cm, yshift=+0.3cm] {\textcolor{orange}{\Large{0.3\%}}} ;
        \node (l0-01) [below of=l0-03, xshift=+0.15cm, yshift=+0.0cm] {\textcolor{olive}{\Large{0.1\%}}} ;

        \node (loo) [above right of=panel, yshift=+5.5cm, xshift=+2.8cm] {\Large{$l_\Gamma \rightarrow \infty$}} ;
        \node (loo-10) [below of=loo, xshift=-0.30cm, yshift=-8.5cm] {\textcolor{darkgray}{\Large{10\%}}} ;
        \node (loo-3) [below of=loo-10, xshift=+0.80cm, yshift=+0.7cm] {\textcolor{blue}{\Large{3\%}}} ;
        \node (loo-1) [below of=loo-3, xshift=+0.80cm, yshift=+0.8cm] {\textcolor{teal}{\Large{1\%}}} ;
        \node (loo-03) [below of=loo-1, xshift=+0.90cm, yshift=+0.8cm] {\textcolor{cyan}{\Large{0.3\%}}} ;
        \node (loo-01) [below of=loo-03, xshift=+0.70cm, yshift=+0.6cm] {\textcolor{olive}{\Large{0.1\%}}} ;
    \end{tikzpicture}
    \caption{
        (\emph{top}) Estimates of uncertainty in $\tilde\Lambda$ as a function of S/N for Cosmic Explorer~\cite{CEpsd} with various levels of calibration uncertainty, and (\emph{bottom}) the ratio of $\sigma_{\tilde\Lambda}$ including calibration uncertainty to $\sigma_{\tilde\Lambda}$ that assumes perfect calibration.
        I show the limiting cases in which (\emph{left}) $\Gamma$ is diagonal ($l_\Gamma=0$, also shown in Fig.~\ref{fig:tides}) and (\emph{right}) $l_\Gamma\rightarrow\infty$ to contrast the limiting behavior for large S/N.
        When $\Gamma$ is diagonal, I show (\emph{dark to light}) $\sigma_\Gamma \sqrt{\Delta f} = 10\%$, 3\%, 1\%, 0.3\%, and 0.1\%.
        When $l_\Gamma\rightarrow\infty$, I show (\emph{dark to light}) $\sigma_\Gamma = 10\%$, 3\%, 1\%, 0.3\%, and 0.1\%.
    }
    \label{fig:tides big l}
\end{figure}

Taking this to an extreme, the scaling of the Fisher information matrix is fundamentally different for finite S/N and $\sigma_\Gamma$ when one first takes the limit $l_\Gamma \rightarrow \infty$.
As discussed in Sec.~\ref{sec:big correlations}, the correction to $C^{-1}$ within $B^{-1}$ scales as $\sigma_\Gamma^2\rho^2/(1+\sigma_\Gamma^2\rho^2)$.
This implies that, when the product $\sigma_\Gamma \rho$ is large, the correction ends up independent of both.
That is, $\lim_{\rho\rightarrow\infty} B^{-1} \rightarrow \epsilon C^{-1}$, which implies that $I_{ab} \approx \partial_a h^\mathrm{T} B^{-1} \partial_b h$ will scale with $\rho^2$ both at low and high S/N.
Indeed, Fig.~\ref{fig:tides big l} shows this behavior explicitly (albeit neglecting the trace term), and calibration uncertainty only ever increases the uncertainty in $\tilde\Lambda$ by at most a factor of a few.

This is in stark contrast to the prediction of Secs.~\ref{sec:big deviations} and~\ref{sec:fisher with large cal uncertainty}, which suggest the uncertainty at high S/N asymptotes to a value that is independent of the S/N.
However, there is no contradiction.
$\Gamma$ becomes singular when $l_\Gamma\rightarrow\infty$, and therefore $\Gamma^{-1}$ does not exist.
As such, the expressions in Secs.~\ref{sec:big deviations} and~\ref{sec:fisher with large cal uncertainty}, which rely
 on $\Gamma^{-1}$, do not apply.

Nonetheless, one may conclude that correlations between frequencies in the calibration prior help reduce the amount of information lost within the astrophysical inference.
However, perfect calibration between frequencies is unlikely to ever be realized in practice, and these conclusions depend to some extent on the assumption of independent amplitude and phase errors as well as the particular assumption of a squared-exponential GP.
Further work may be needed to determine whether GPs based on more realistic calibration measurements also show this behavior.

\section{On the interpretation of $\sigma_\Gamma \sqrt{\Delta f}$}
\label{sec:interpretation}

The careful reader may have been struck by the apparent dependence of my results on the frequency spacing within, e.g., Eq.~\ref{eq:fisher l=0}, where the relevant quantity describing the calibration uncertainty is $\sigma_\Gamma^2 \Delta f$.
While this should not be surprising, as the variance of the data due to stationary Gaussian noise has a similar scaling ($C\Delta f \sim s$ is constant for different $\Delta f$), it is perhaps illuminating to consider this in more detail.

Specifically, consider downsampling the data by averaging the data in neighboring frequency bins.
\begin{equation}
    d_{i+j} = \frac{1}{2}(d_i + d_j)
\end{equation}
Applying this procedure to the signal model, one obtains
\begin{align}
    \frac{1}{2}(h_i(1+\delta_i) + h_j(1+\delta_j))
        & = \frac{1}{2}(h_i + h_j) + \frac{1}{2}(h_i \delta_i + h_j \delta_j) \nonumber \\
        & = h_{i+j}\left( 1 + \frac{h_i \delta_i + h_j \delta_j}{h_i + h_j} \right)
\end{align}
from which it is clear that $\delta_{i+j}$ is a signal-weighted average of $\delta_i$ and $\delta_j$.
Thus
\begin{equation}
    \mathrm{Var}(\delta_{i+j}) = \frac{h_i^2}{(h_i+h_j)^2} \mathrm{Var}(\delta_i) + \frac{h_j^2}{(h_i+h_j)^2} \mathrm{Var}(\delta_j) + 2 \frac{h_i h_j}{(h_i+h_j)^2} \mathrm{Cov}(\delta_i, \delta_j)
\end{equation}
If $\mathrm{Var}(\delta_i) = \mathrm{Var}(\delta_j) = \mathrm{Cov}(\delta_i, \delta_j) = \sigma_\Gamma^2$, implying $\delta_i$ and $\delta_j$ are strongly correlated, then $\mathrm{Var}(\delta_{i+j}) = \sigma_\Gamma^2$.
However, if $\delta_i$ and $\delta_j$ are uncorrelated ($\mathrm{Cov}(\delta_i, \delta_j)=0$), then $\mathrm{Var}(\delta_{i+j}) = \sigma_\Gamma^2 (h_i^2 + h_j^2)/(h_i+h_j)^2$.
Furthermore, if the signal does not change much between neighboring frequency bins, then $\mathrm{Var}(\delta_{i+j}) \approx \sigma_\Gamma^2 / 2$.
That is, the variance of the averaged $\delta$ is smaller than the original variance by a factor that is proportional to the number of frequency bins that were averaged.

Therefore, for diagonal $\Gamma$, one might expect the product of $\Gamma$ and the frequency spacing to be constant, as the frequency spacing scales with the number of samples that have been averaged together.
If diagonal $\Gamma$ are realizable in practice, it will likely be better to describe them in terms of $\Gamma \Delta f \sim \mathrm{constant}$ just as $C \Delta f \sim s$ is a constant.

\end{document}